\documentclass[10pt]{article}

\usepackage[superscript,sort]{cite}

\usepackage[pdftex,colorlinks=true,linkcolor=black,citecolor=black,urlcolor=black,bookmarks=true,breaklinks=true]{hyperref}
\usepackage[pdftex]{graphicx}
\usepackage[usenames,pdftex]{color}
\definecolor{Red}{rgb}{1.0,0.0,0.0}

\usepackage{amsthm,amscd,amsxtra,amsfonts,amsmath,amssymb,multirow}
\usepackage{wrapfig}
\usepackage[footnotesize]{caption}
\usepackage[tiny,compact]{titlesec}

\usepackage{natbib}
\usepackage{subfigure}
\usepackage{tikz}
\usetikzlibrary{shapes,arrows}
\usepackage{ctable}

\titlespacing*{\section}{0pt}{*0}{*0}
\titlespacing*{\subsection}{0pt}{*0}{*0}
\titlespacing*{\subsubsection}{0pt}{*0}{*0}
\titlespacing{\paragraph}{0pt}{*0}{*1}

\definecolor{MyPurple}{rgb}{1,0,1}

\newcommand{\beq}[1]{\begin{equation} \label{#1}}
\newcommand{\eeq}{\end{equation}}
\newcommand{\barray}{\begin{array}{ll}}
\newcommand{\earray}{\end{array}}

\begin{document}

\pagenumbering{roman}

\clearpage \pagebreak \setcounter{page}{1}
\renewcommand{\thepage}{{\arabic{page}}}

\title{A Review of Mathematical Modeling, Simulation and Analysis of Membrane Channel Charge Transport  }

\author{
Duan Chen$^{1}$\footnote{E-mail: Duan.Chen@uncc.edu} ~ and
Guo-Wei Wei$^{2,3}$\footnote{E-mail: Wei@math.msu.edu},
\\
$^1$Department of Mathematics and Statistics\\
University of North Carolina at Charlotte, NC 28223, USA\\
$^2$Department of Mathematics \\
Michigan State University, MI 48824, USA\\
$^3$Department of Biochemistry  and Molecular Biology \\
Michigan State University, MI 48824, USA \\
}

\date{\today}
\maketitle

\begin{abstract}
The molecular mechanism of ion channel gating and substrate modulation is elusive for many voltage gated ion channels, such as  eukaryotic  sodium ones. The  understanding of channel functions is a pressing issue  in molecular biophysics and biology. Mathematical modeling, computation and analysis of  membrane channel charge transport have become an emergent field and give rise to significant  contributions to our understanding of ion channel gating and function.  This review  summarizes recent progresses and outlines remaining challenges in mathematical modeling, simulation and analysis of ion channel charge transport. One of our focuses is the	Poisson-Nernst-Planck (PNP) model and its generalizations.  Specifically,  the basic framework of the PNP system and some of its extensions, including size effects, ion-water interactions,  coupling with density functional theory and relation to fluid flow models.  A reduced theory, the Poisson-Boltzmann-Nernst-Planck (PBNP) model, and a differential geometry based ion transport model are also discussed. For  proton channel, a multiscale and multiphysics Poisson-Boltzmann-Kohn-Sham (PBKS) model is presented.   We show that all of these ion channel models can be cast into a unified  variational multiscale framework with a macroscopic continuum domain of the solvent and a microscopic discrete domain of the solute. The main strategy is to construct a total energy functional of a charge transport system to encompass the polar and nonpolar free energies of solvation and chemical potential related energies. Using the Euler-Lagrange variation, the coupled PNP equations and other transport equations are derived, whose solutions lead to the minimization of the total free energy and explicit profiles of electrostatic potential and densities of charge species. 
Current computational algorithms and tools for numerical simulations and results from mathematical analysis of ion channel systems are also surveyed.
\end{abstract}

{\bf Key words:}
Variational multiscale models,
Charge transport,
Ion channels,
Laplace-Beltrami equation,
Poisson-Boltzmann equation,
Nernst-Planck equation,

\newpage

{\setcounter{tocdepth}{4} \tableofcontents}

\vskip 12pt

\section{Introduction}\label{sec:int}

Membrane charge transport is one of the most important biological processes in lives, as it facilitates signal transduction, action potential,  cardiac rhythms, muscle contraction, T-cell activation, etc.  
Transporting charges could be small proteins, mobile ions, and dipoles in solvent environment, and they are conducted by membrane  channels such as ion channel, proton pumps, or general transporter proteins. In a specific but critical manner, ion channels are a type of transporter proteins embedded in cell membranes.  They have tube-like water pores  in the middle that facilitate selected ion permeation and maintain proper cellular ion compositions.  Phospholipid bilayer of cells provides a  hydrophobic barrier to the passage of charged ions in extra- and intra-cellular environment, but strongly polar or even charged amino acids of channel proteins provide a conducting pathway across the hydrophobic interior of the membrane bilayer  (\cite{Ikezu:2008,Karniadakis:2005}).     
	Resulting ionic flux plays a key role in almost many physiological phenomena from nerve and muscle excitation, human sensory transduction, to cell volume and blood pressure regulation, etc. More critically,   ion channels are prominent factors to human health. One example is in cancer research: glioblastoma multiforme exhibits abnormal upregulation of gBK potassium ( K$^+$)  and ClC-3 chloride (Cl$^-$) channels, which aid glioblastoma cells changing cellular volume very rapidly, thus help extremely aggressive invasive behavior of the tumor cells. Another example is the M2 proton (H$^+$) channel in influenza A virus. The M2 proton channel conducts protons into the virion core, acidifies the virus interior, and leads  viral ribo nucleo protein (RNP) complexes release and start viral replication.  It has been intensively studied that ion channels are frequent targets in research of new drugs for human diseases (\cite{Fermini:2008,YMiao:2015,HDong:2013,TACross:2012,HXZhou:2011}). Advanced giga-seal parch-clamp technique has made the measurement of ionic flowing through a single channel possible  (\cite{Chung:2002}), and experimental discoveries are available for structures and functions of potassium channels (\cite{Bezanilla:2015, MacKinnon:2014}), sodium channels  (\cite{Catterall:2012, Payandeh:2012}), calcium channels  (\cite{Catterall:2014}), and proton channels  (\cite{DeCoursey:2014}). These developments set the stage for theoretical/mathematical modeling to simulate charge transport, to reproduce experimental data,  to predict new phenomena, and to offer direction in ion channel-targeted drug design  (\cite{Boiteux:2014,Lacroix:2013}).

\begin{figure}
            \begin{center}
                \begin{tabular}{cc}
                         \includegraphics[width=0.5\columnwidth]{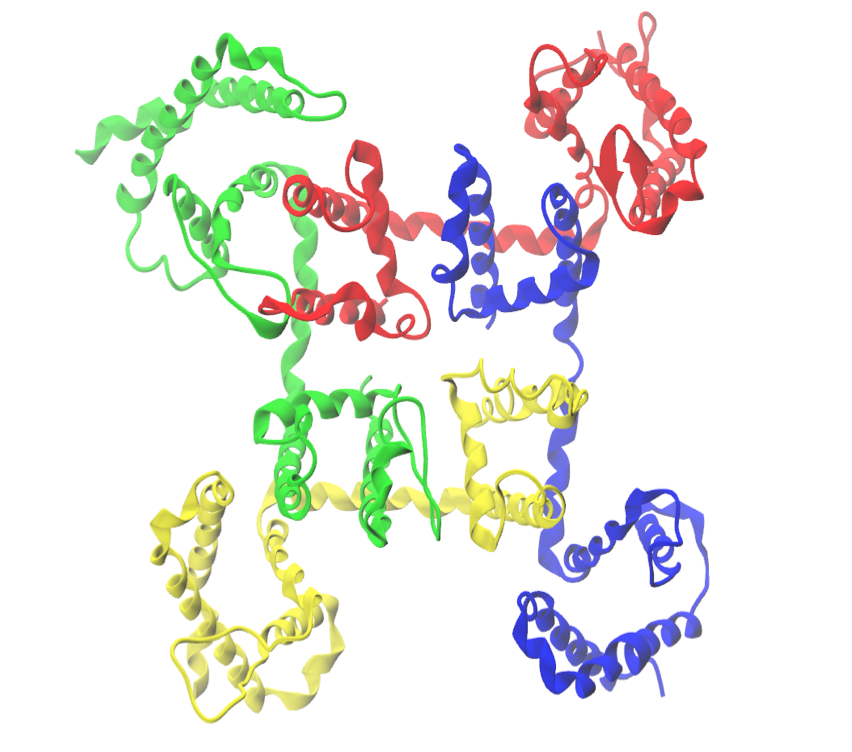}&
                    \includegraphics[width=0.5\columnwidth]{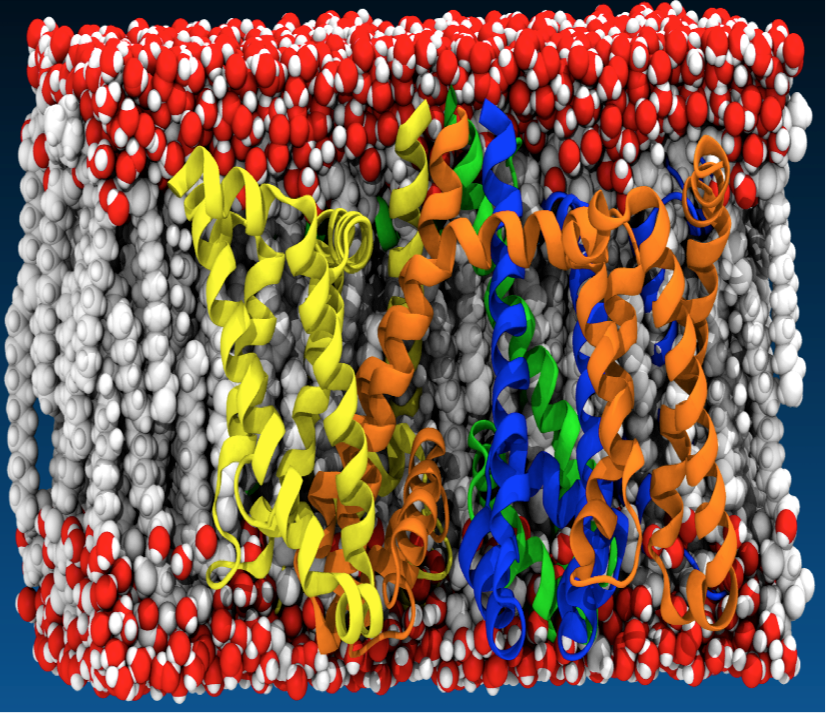}\\
                    (a)&(b)\\
                \end{tabular}
            \end{center}
             \caption{Homology structure of a sodium channel constructed from a mosquito genetic data.
             (a)   Top view of a  mosquito voltage gated sodium channel. Four homologous  domains 	(I through IV) 	are illustrated by using four different colors. In each domain, there are  six transmembrane segments (S1 through S6). 	Segments S1$-$S4 of the channel constitute the voltage-sensing domain (VSD), which is away from the pore, while segments S5 and S6, together with the membrane-reentrant pore loop, form the pore.  Image credit: Zixuan Cang.
             (b) The sodium channel  viewed in a cell membrane. Image credit: Christopher Opron.  }
             \label{fig:nachannel}
\end{figure}

		There is quite good understanding of the molecular basis of voltage gated potassium channels across various species, such as bacteria, insects, and mammals, due to the availability of many high quality X-ray crystallographic structures (\cite{Gutman:2005,SBLong:2005}), and detailed theoretical analysis (\cite{Jensen:2012}).  Voltage gated potassium channels typically are simple, single-domain proteins that assemble to form functional homotetramers.  However, eukaryotic voltage gated sodium channel (Nav) channels, such as those of mosquitoes and humen, are complex and four homologous domain proteins  assembled into pseudotetrameric structures, with no X-ray crystallographic structure, see Fig. \ref{fig:nachannel}.  The only existing high resolution X-ray crystallographic structures for Nav channels are from bacteria (\cite{McCuskeretal:2012,Bagnerisetal:2013,Shayaetal:2014,Payandeh:2012}).   Similarly to voltage gated potassium channels, bacterial Nav channels have four identical subunits arranged to form a functional channel. The molecular mechanism of eukaryotic  Nav channel gating transitions, between voltage-sensing, activation, and deactivation as well as their coupling and interaction with drugs and/or insecticides, is largely unknown or unclear, partially due to limited experimental means, theoretical models and computational power  (\cite{Duetal:2013, Dongetal:2014, Duetal:2015}). Such a gap in our understanding severely hinders our ability to design effective mosquito insecticides as well as qualified  drugs for epilepsy, irregular cardiac arrhythmias, hyperalgesia, myotonia, and anesthesia.

\begin{figure}
            \begin{center}
                \begin{tabular}{cc}
                         \includegraphics[width=0.5\columnwidth]{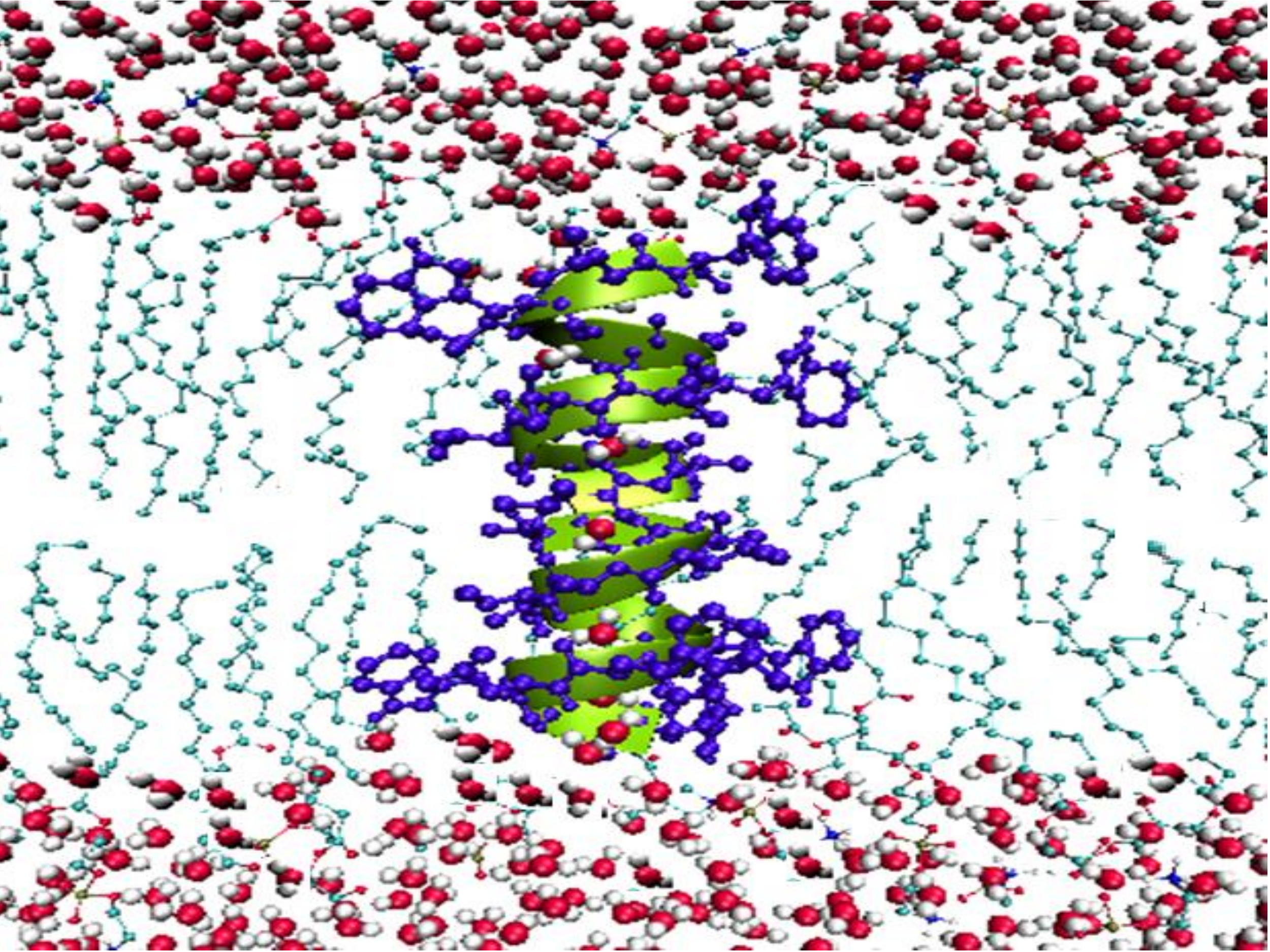}&
                    \includegraphics[width=0.5\columnwidth]{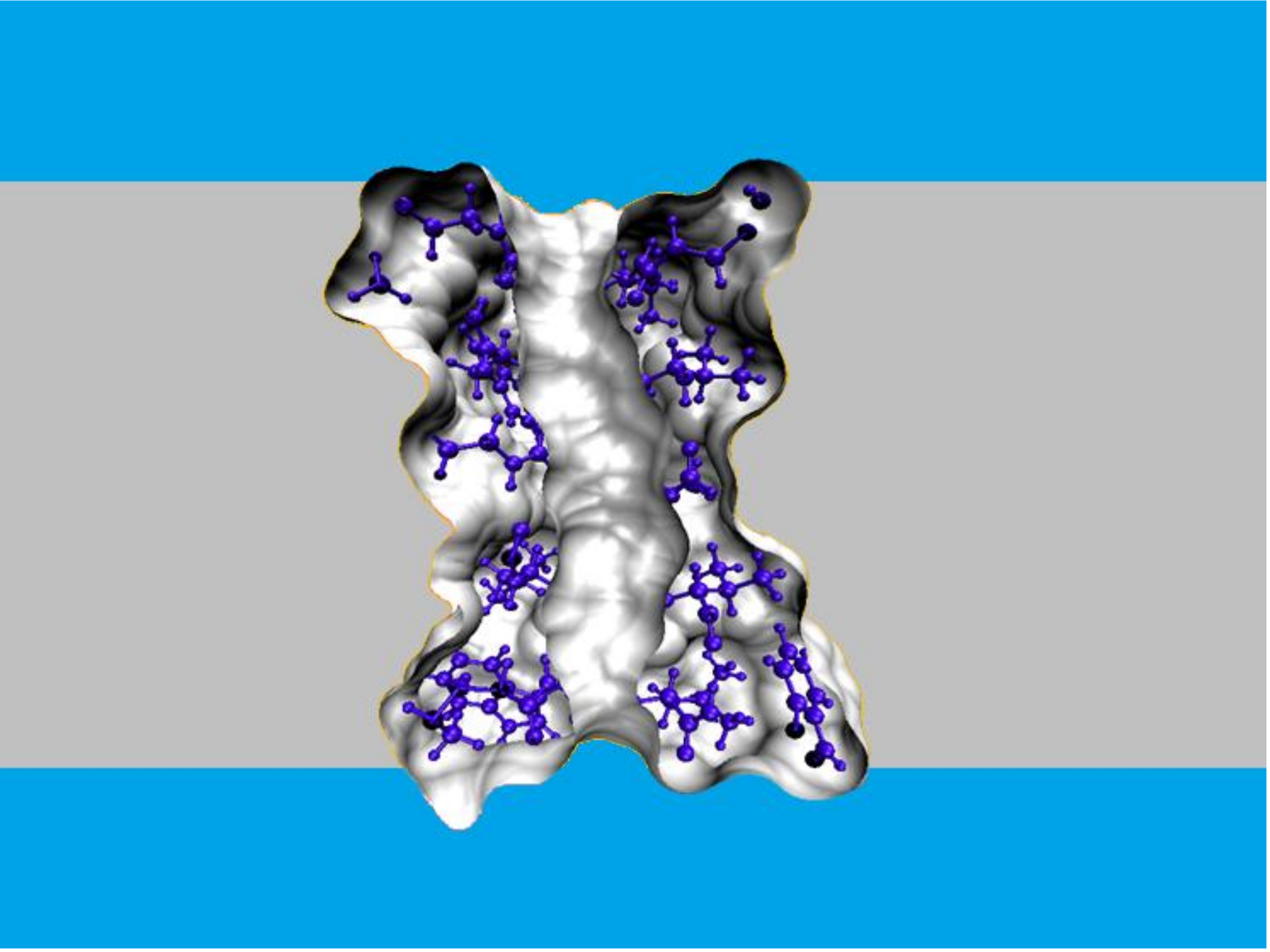}\\
                    (a)&(b)\\
                \end{tabular}
            \end{center}
             \caption{Illustration of an ion channel system and the multiscale approach.
             (a) Atomic view of the Gramicidin A channel in the membrane and aqueous environment;
             (b) A cross section of the multiscale representation of the system. }
             \label{fig1magcomplex}
\end{figure}

Modern ion channel models are based on the biological understanding of the function and gating mechanism of ion channels, which heavily depend on their molecular structures. A channel protein usually consists of several hundreds to thousands amino-acid residues. Some of them have simpler structures, for example,  the Gramicidin A (GA) channel obtained from the soil bacterial species Bacillus brevis  is   just a dimer that consists of two head-to-head $\beta$-helical parts in a bilayer membrane. All of the residues form a narrow pore of about 4\AA~in diameter and 25\AA~ in length that simply conducts monovalent cations, binds bivalent cations, while rejects anions.  
	On the contrast,  the  KcsA  (potassium crystallographically-sited activation) channel has a relatively complicated structure and hence complete functions. It is comprised of around 560 residues that form four identical subunits, each containing two alpha-helices connected by a loop of approximately 30 amino acids. There are three primary functioning sections of the KcsA channel: the opening pore on the cytoplasmic side of the cell interior, a small cavity filled with water and a mix of sodium ($\mathrm{Na^{+}}$) and potassium ($\mathrm{K^{+}}$) ions, and the selectivity filter  (\cite{Egwolf-Roux-2010}).  The KcsA channel  is specialized to facilitate and regulate
the conduction of $\mathrm{K^{+}}$ ions in particular  (\cite{UofIwebsite,Egwolf-Roux-2010}). Further, the technology of homology provides ability to discover more ion channel three dimensional (3D) structures from their genetic data. Figure \ref{fig:nachannel} shows the homology structure of a sodium channel constructed from a mosquito genetic data.  In fact, voltage gated potassium and sodium channels are homologous to each other. Therefore, a full description of of ion channels at {\em atomic level} is required in order to investigate their function and gating  comprehensively.
	On the other hand, a major characteristics of an ion channel system is its heterogeneity. It is inhomogeneous in terms of materials: as shown in Fig. \ref{fig1magcomplex}(a), an ion channel exists in an extremely complicated environment including cell membrane, water molecules, mobile ions and other molecular components. All of these components are subject to intensive mutual long-range (e.g., electrostatics) and short-range (e.g. Lennard-Jones) interactions. It is also inhomogeneous in terms of functions:  an ion channel may have both water-rich, or bulk solvent regions where ions are fully hydrated, and a  region where water molecules can only form a single file and ion-water clusters need to be rearranged. Additionally, an ion channel may have different function domains that are responsible for voltage sensing, gating, conducting efficiency and selectivity. In fact, some of these domains may extend away from the channel pore, as in the case of voltage gated sodium channels  (\cite{Catterall:2012, Payandeh:2012}). Furthermore, ion channel gating can be modulated by ligand,  toxicity, substrate, et cetera, which is often the basis  for  anesthesia and insect control.	Therefore, theoretical modeling and simulations of the molecular mechanism for charge transport in ion channels often involve an excessively large number of degrees of freedom and encounter enormous challenges   (\cite{Wei:2009,Wei:2012,Wei:2013}).

	One of the pioneering works about the functions of  ion channels is the Hodgkin-Huxley model  derived by \cite{HH:1952}, which is used to study action potentials initiated and propagated in neurons by modeling  the ensemble of voltage dependent channels in nerve fibers using nonlinear ordinary differential equations. This model is still a basis for many present interesting ion channel studies.  
		In this review, we focus on models for the functions of a single ion channel. Among mathematical/theoretical models of ion channels developed in past decades, {\em molecular dynamics} (MD) stands  among top of the hierarchy in terms of accuracy. As a type of all-atom channel model, it treats  channel protein, cell membrane, solvent and ions explicitly  with positions and velocities propagated by the Newtonian dynamics.   MD simulation provides a way to investigate the ultimate details of how structures move and which motions may be linked to biological functions. From an MD simulation trajectory, a variety of thermodynamic (e.g., the free energy changes associated with solvation or ion binding inside a channel pore) or kinetic (e.g., the rate of ion passing through a channel) quantities can be calculated. Computer tools based on MD have been extensively developed by  \cite{AMBER, CHARMM22,Klapper:1986,Prabhu:2008,UHBD,MEAD} and simulations have been widely used to compute energetics (potential of mean forces PMFs) of ion or water transport through biological channels. Further, MD is a suitable model to study some specific function unit of ion channels such as selectivity mechanism in the local selectivity filters. However, the major drawback of the explicit method is the extremely large number of degree of freedom for the system, so the computation remains very expensive even with contemporary computer powers due to the necessarily small time step ($10^{-15}$ seconds) versus the ion permeation time scale ($10^{-6}$ seconds). Some important ion channel models in MD are referred to   \cite{Marx:2000,Roux:2004,Im:2002,Roux1:2002,Schumaker:2000, ZhouHX:2013, ZhouHX:2014, Caiw:2013, Caiw:2016}. Specifically, gating motion of ion channels, and their interactions between transmembrane were studied in   \cite{ShawD:2012, ZhouHX:2013, ZhouHX:2014,ZhouHX:2015}.
	
	{\em Brownian dynamics} (BD) fades out the molecular details of membrane bilayer and water molecules, while only treats mobile ions and channel protein explicitly. The motion of target ions is governed by the Langevin equation. In this model, the forces acting on the ions include frictional/random forces from the surrounding solvent, and the total electrostatic forces due to other mobile ions, fixed charges in the channel protein, solvent polarized electrostatic field and/or applied transmembrane potential that are determined by solving the Poisson equation. Many BD based algorithms are developed and because of the significant reduction in the number of degrees of freedom. The BD simulations are highly efficient. The trajectories of mobile ions can be simulated at usually microsecond scale, from which the single channel current can be derived by counting the number of ions that move across the channel. A series of work has been established to study functions of channel proteins via BD  by \cite{Cheng:2010,Gordon:2009,Cheng:2005,Coalson:2005}.
	
	One of simplest  ion channel models is the {\em Poisson-Nernst-Planck} (PNP) theory, which is a mean-field approach with a low resolution of ion channels, but offers  high efficiency.  Using a continuum approximation, the PNP model treats the ion flow as the averaged ion concentration driven by the electrostatic potential force and ion concentration gradient.  Meanwhile, it incorporates the static atomistic charge description of channel proteins. Thus, it hybrids the macroscopic/continuum description of ionic channel flows with the microscopic/discrete representation of protein electrostatic charge sources, see Fig \ref{fig1magcomplex}(b). For a 1:1 electrolyte, the PNP system (without physical parameters) is essentially a system of coupled partial differential equations (PDEs):
	\begin{equation}\label{eqn:oldpnp}
		\left\{\begin{array}{ll}
			-\epsilon\Delta\phi=\rho_f+p-n, \\
			\\
			\displaystyle{\frac{\partial p}{\partial t}}=\nabla\cdot\left(\nabla p+p\nabla\phi\right) \\
			\\
			\displaystyle{\frac{\partial n}{\partial t}}=\nabla\cdot \left(\nabla n-n\nabla\phi\right),
		\end{array}\right.
	\end{equation}
	 where $\phi({\bf r})$, $p({\bf r}, t)$ and $n({\bf r}, t)$ are electrostatics, concentrations of positive mobile ions (cations) and negative mobile ions (anions), receptively.  In this approach, both water and cell membrane are approximated by dielectric function $\epsilon$, and the structure of channel protein is modeled by  static point charges or atomic charge  density $\rho_f$. Under this framework, concentrations of the ions through the channel follow the Ohm's and Fick's law, and form two drift-diffusion equations in the same structure. 
	 The PNP model was introduced to the field of molecular biology in  early 1990s    by \cite{Eisenberg:1993,DPChen:1995,Barcilon:1992,Singer:2008,Eisenberg:1996,Eisenberg:2010} from a similar approach, called drift-diffusion  equations in electronic devices community and widely used in ion channel simulations afterwards, as one of the current major workhorses  (\cite{Hwang:2006,Dyrka:2008,Cardenas:2000,Singer:2008,Chen:1997,Kurnikova:1999,Corry:2003,Gillespie:2002,Allen:2001,Choudhary:2010,Constantin:2007,Graf:2004, Levitt:1999,Mamonov:2006, Simakov:2010}).

	 But the reality of ion channels is far more than three PDEs. 
	 The starting point of the fundamental PNP system in Eq. (\ref{eqn:oldpnp}) is the assumption of volume-less point charge approximation of mobile ions embedded in a structureless and homogeneous continuum model of water molecules. Thus, the original model neglects the steric effects of ions in significant geometric confinement in channel pore, ion-water interactions, polarization of water molecules, ion-ion correlation and fluctuations, channel motion, and many other interactions that are not directly and exclusively relate to electrostatics, but critically contribute to complicated functions of ion channels such as selectivity and activation.	Some theoretical approaches have been proposed towards the direction of improving the traditional Poisson-Boltzmann or PNP theory, such as in  \cite{Ben-Yaakov:2011b, Antypov:2005,Bazant:2005, Bazant:2011, Vlachy:1999,Wei:2009,Wei:2012}. Development, analysis, and computation of PNP-based models for ion channel transport have also attracted much attention in the community of applied mathematics.


In this review, we first review the fundamental PNP model and its application on a simple ion channel in Section \ref{sec:PNP}. The recent progresses on new models development, computational methods, and mathematical analysis are surveyed in Sections \ref{sec:generalizedPNP}, \ref{sec:algorithm}, and \ref{sec:analysis}, respectively. Finally, we discuss some other advanced mathematical models in Section \ref{sec:others},  including  the Poisson-Boltzmann-Kohn-Sham model  for proton channels and a differential geometry based multiscale model for a comprehensive understanding of solvation and charge transport.



	\section{The  Poisson-Nernst-Planck model}\label{sec:PNP}
	
%
		The variational derivation of Poisson-Boltzmann (PB) equation was given in early 1990s   by \cite{Sharp:1990}. Similarly, the system of PNP equations can be derived by the variation of a total energy functional  (\cite{Fogolari:1997}). However, unlike the PB equation or the Poisson equation, which can derived entirely from total variation, the derivation of the Nernst-Planck equation follows two steps, namely, using the energy variation to obtain the chemical potential and then  using Fick's laws of diffusion to attain the Nernst-Planck equation. A somewhat more rigorous derivation from the conservation laws has been given recently  by \cite{Wei:2009}. This approach allows the coupling to flow velocity and potential chemical reactions, such as  protonation and deprotonation of amino acids, which occur very often during ion channel   permeation.

In this review, we follow a simple derivation  	as in	  \cite{Fogolari:1997, QZheng:2011b}.	
		Assuming there are multiple ionic species in an ion channel system and $\rho_{\alpha}$ is the concentration of the $\alpha$-th ion species, the total free energy for the system can be described in terms of the electrostatic potential $\Phi$ and the concentration $\rho_{\alpha}$  as the following:
\begin{eqnarray}
\begin{aligned}
G^{\rm PNP}_{\rm{total}}[\Phi,\{\rho_\alpha\}] &= \int\left\{-\frac{\epsilon_m}{2}|\nabla\Phi|^2 + \Phi \ \rho_m-\frac{\epsilon_s}{2}|\nabla\Phi|^2+\Phi\sum_{\alpha} \rho_{\alpha}q_{\alpha}\right. \\\label{eq17tot}
&+\left.\sum_\alpha\left[ \left(\mu^0_{\alpha }-\mu_{\alpha 0} \right) \rho_\alpha + k_B T  \rho_\alpha  {\rm{ln}} \ \frac{\rho_\alpha }{ \rho_{\alpha 0} } - k_B T \left(\rho_\alpha  - \rho_{\alpha 0} \right)  + \lambda_\alpha \rho_\alpha \right]\right\}d{\bf{r}}.
\end{aligned}
\end{eqnarray}

The first row of Eq. (\ref{eq17tot}) is the electrostatic free energies of the system.  Here, the ion channel protein is modeled as the fixed charge density $\rho_m$ in atomic details with dielectric constant $\epsilon_m$. In contrast, the solvent is modeled by the ionic density $\rho_{\alpha}$ and water molecules are treated as a dielectric continuum with dielectric constant $\epsilon_s$. Here $q_\alpha$ is the charge of $\alpha$th ion species.  
The second row includes chemical potential related energy and entropy of mobile ions, where $\mu_{\alpha}^0$ is the reference chemical potential of the $\alpha$th species at which the associated reference concentration is $\rho_{\alpha 0}$ and  $k_BT$ is the thermal energy with $k_B$ being the Boltzmann constant and $T$ being the temperature. 
At last,  a Lagrange multiplier $\lambda_\alpha$ is used  to ensure appropriate physical properties at equilibrium   (\cite{Fogolari:1997}).

By applying the  variational principle, governing equations for the variables $\Phi$ and $\rho_{\alpha}$ of the system can be obtained.

\subsection{Governing equations}

First, the Poisson equation can be derived   by taking the variation with respect to the electrostatic potential $\Phi$, i.e, 
\begin{eqnarray}\label{eq19varphi}
\frac{\delta G^{\rm PNP}_{\rm{total}}}{\delta \Phi} \Rightarrow
    \nabla\cdot\left(\epsilon \nabla\Phi \right)+ \rho_m
    +\sum_{\alpha} \rho_{\alpha}q_{\alpha}=0.
\end{eqnarray}
Then it yields
\begin{eqnarray}\label{eq24poisson}
-\nabla\cdot\left(\epsilon \nabla\Phi \right)= \rho_m
    +\sum_{\alpha} \rho_{\alpha}q_{\alpha}.
\end{eqnarray}
In many multiscale models such as in \cite{DuanChen:2016c,QZheng:2011a},  the Poisson equation is defined in the whole computational domain $\Omega$, which consists of the solute domain $\Omega^+$ and the solvent domain $\Omega^-$, on which the dielectric function $\epsilon$ is defined as a piecewise constant function
	\begin{eqnarray}\label{eqn:epsilon}
\displaystyle  \epsilon= \left\{\begin{aligned}
    & \epsilon_m,
    & \mathbf{r} \in \Omega^+, \\
    &\epsilon_s,
    & \mathbf{r} \in\Omega^-
 \end{aligned}\right.
\end{eqnarray}
At this moment domains $\Omega^+$ and $\Omega^-$ are assumed to be divided by a given molecular surface $\Gamma$, i.e., $\Omega=\Omega^+\cup\Omega^-$ and $\Gamma=\Omega^+\cap \Omega^-$.

The derivation of  the Nernst-Planck equation follows two steps. First, consider the variation of the total free energy functional with respect to ion concentration  $\rho_{\alpha}$:
\begin{equation}\label{eq20varn}
\frac{\delta G^{\rm PNP}_{\rm{total}}}{\delta \rho_\alpha} \Rightarrow
\mu^{\rm gen}_\alpha= \mu^0_{\alpha }-\mu_{\alpha 0} + k_B T  {\rm{ln}} \ \frac{\rho_\alpha }{ \rho_{\alpha 0}} + q_{\alpha} \Phi + \lambda_\alpha
=\mu^{\rm chem}_\alpha + q_{\alpha} \Phi  +\lambda_\alpha,
\end{equation}
where $\mu^{\rm gen}_\alpha$ is the relative generalized potential of species $\alpha$ and it vanishes at  system equilibrium. Hence, one has
\begin{equation}\label{eq20Equil}
\lambda_\alpha=-\mu^0_{\alpha } \quad {\rm, }\quad
\rho_\alpha =\rho_{\alpha 0}e^{-\frac{q_{\alpha }\Phi -\mu_{\alpha0}}{k_B T }},
\end{equation}
and therefore
\begin{equation}\label{eq21mu}
\mu^{\rm gen}_\alpha=
 k_B T {\rm{ln}} \ \frac{\rho_\alpha }{ \rho_{ \alpha 0} }\ + q_{\alpha} \Phi   -\mu_{\alpha0}.
\end{equation}
Then, by Fick's first law, the ion flux ${\bf J}_\alpha$ is given through the gradient of the
relative generalized potential, i.e.,  ${\bf J}_\alpha=-D_\alpha \rho_\alpha \nabla \frac{\mu^{\rm gen}_\alpha}{k_B T}$, where
$D_{\alpha}$ is the diffusion coefficient of species $\alpha$.
 If steam velocity and chemical reaction are neglected  (\cite{Wei:2009}), the the mass conservation of species $\alpha$  gives $\frac{\partial \rho_\alpha}{\partial t}=-\nabla \cdot {\bf J}_\alpha$, i.e., 
\begin{eqnarray}\label{eq22nernst}
 \frac{\partial \rho_\alpha}{\partial t}=\nabla \cdot \left[D_{\alpha}
  \left(\nabla \rho_{\alpha}+\frac{ \rho_{\alpha}q_\alpha}{k_{B}T}\nabla \Phi \right)\right],
\end{eqnarray}


%

Equation (\ref{eq22nernst}) is only defined in the solvent domain $\Omega^-$ but   forms a coupled system with Eq. (\ref{eq24poisson})  for describing the charge concentrations $\rho_\alpha$ and the electrostatic potential $\Phi$. The solutions of these equations need to be pursued self-consistently.  Based on the ionic flux, the general formulation to calculate an important physical observable--ionic current,  is
	\begin{equation}\label{current}
		I=\sum_{\alpha =1}^{N_c} q_{\alpha} \int_{L_x,L_y} D_{\alpha} \left( \frac{\partial \rho_{\alpha}}{\partial z} +  \frac{\rho_{\alpha} q_{\alpha}}{k_B T}\frac{\partial \Phi}{\partial z}\right) dx dy.
	\end{equation}

	\subsection{Gramicidin channel: A showcase for the PNP model}
\begin{figure}[ht!]
    \begin{center}
           \begin{tabular}{cc}
         	  \includegraphics[width=0.52\textwidth]{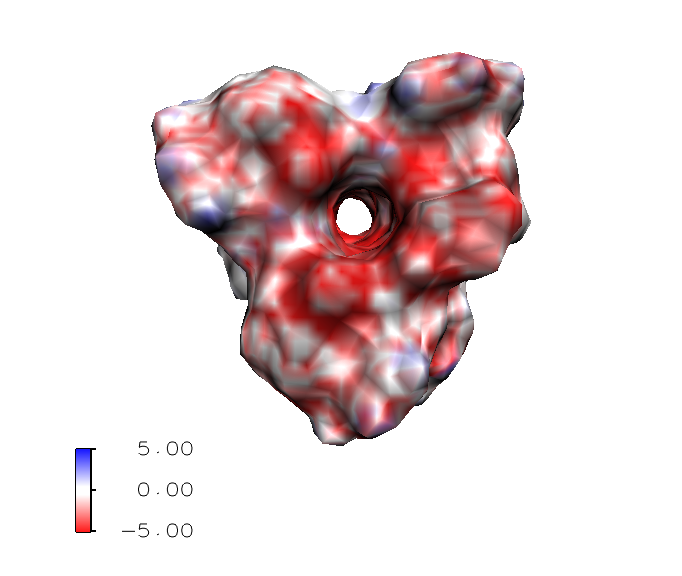}&
          	  \includegraphics[width=0.48\textwidth]{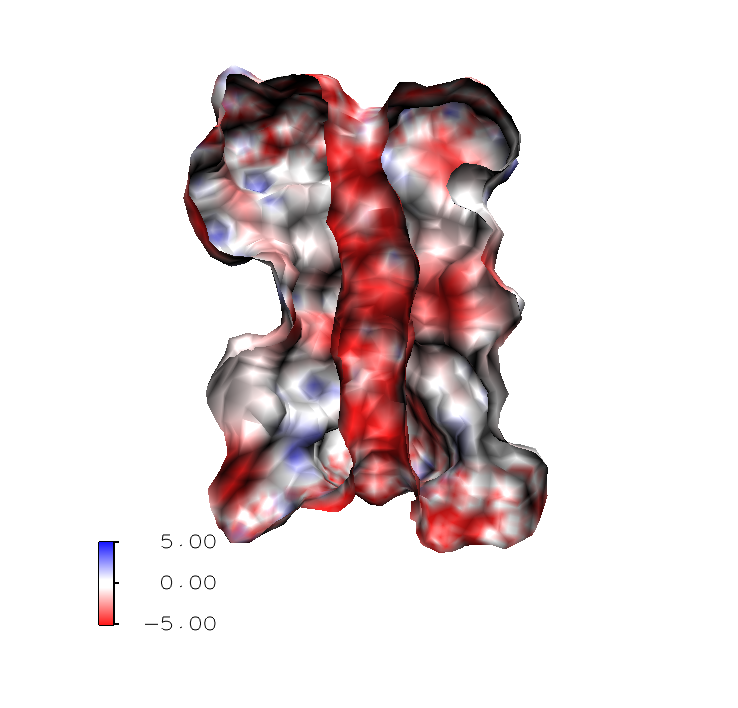}\\
		(a)  & (b) \\
           \end{tabular}
    \end{center}
         \caption{ 3D illustration of the electrostatic profile of the  Gramicidin A (GA) channel. The red and blue colors represent negative and positive electrostatics, respectively. (a) Top view of the GA channel; (b) Side view of the GA channel.}
    \label{fig:1mag-electro}
\end{figure}

	As a showcase, electrostatic profiles of the GA channel, calculated from the PNP system, is mapped on the protein surface and shown in Fig. \ref{fig:1mag-electro}. Overall the GA is neutral in terms of charges, but its surface electrostatic potential  is mostly negative near the channel mouth as indicated by the red color in the graph. Also, the inner wall of the channel pore is also intensively negatively charged as shown in Fig. \ref{fig:1mag-electro}(b).  This fact indicates the obvious selectivity of the GA channel --- it selects cations and suppresses  anions.
	The electrostatics of the channel system greatly depends on the dielectric constants used in Eq. (\ref{eqn:epsilon}).  However, the choice of this key model parameter is very subtle and nontrivial because rotation and polarization of water molecules in the narrow channel pore are significant different from those in bulk solvent. The true dielectric properties of water molecules in ion channel are not fully revealed. To this end,  a range of dielectric constants have been explored  in order to obtain  a reasonable prediction in \cite{DuanChen:2013}. There is a general agreement that $\epsilon_{m}$ could  be taken as a constant that slightly greater than 2, which is the value used in the solvation study. While the dielectric constant $\epsilon_{s}$ for the solvent should be position dependent. The dielectric constant $\epsilon_{\rm bath}=80$ is the value widely accepted in the literature for the bath water region. 
	\begin{figure}[ht!]
            \begin{center}
                \begin{tabular}{cc}
                    \includegraphics[width=0.5\columnwidth]{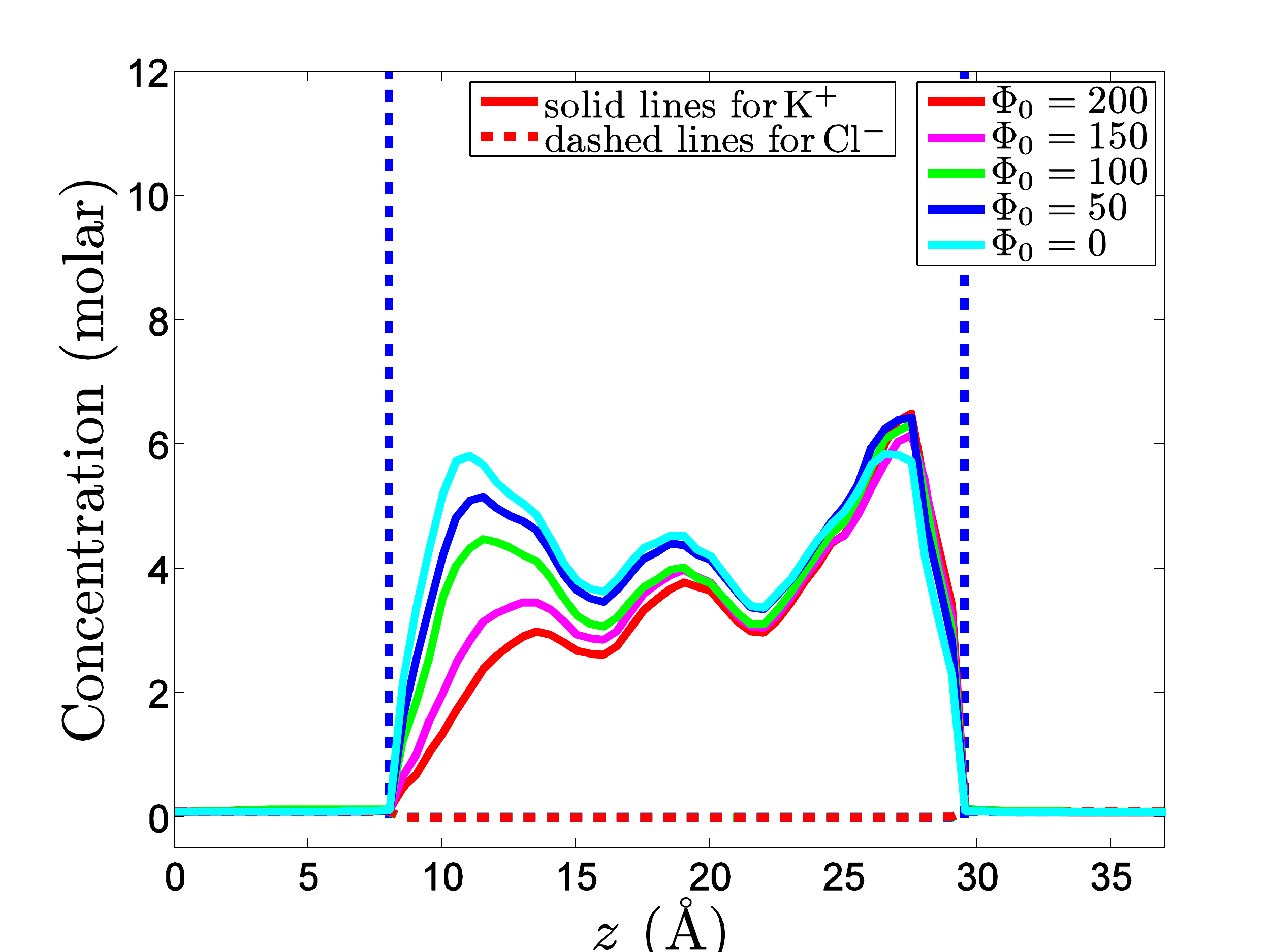}&
                    \includegraphics[width=0.5\columnwidth]{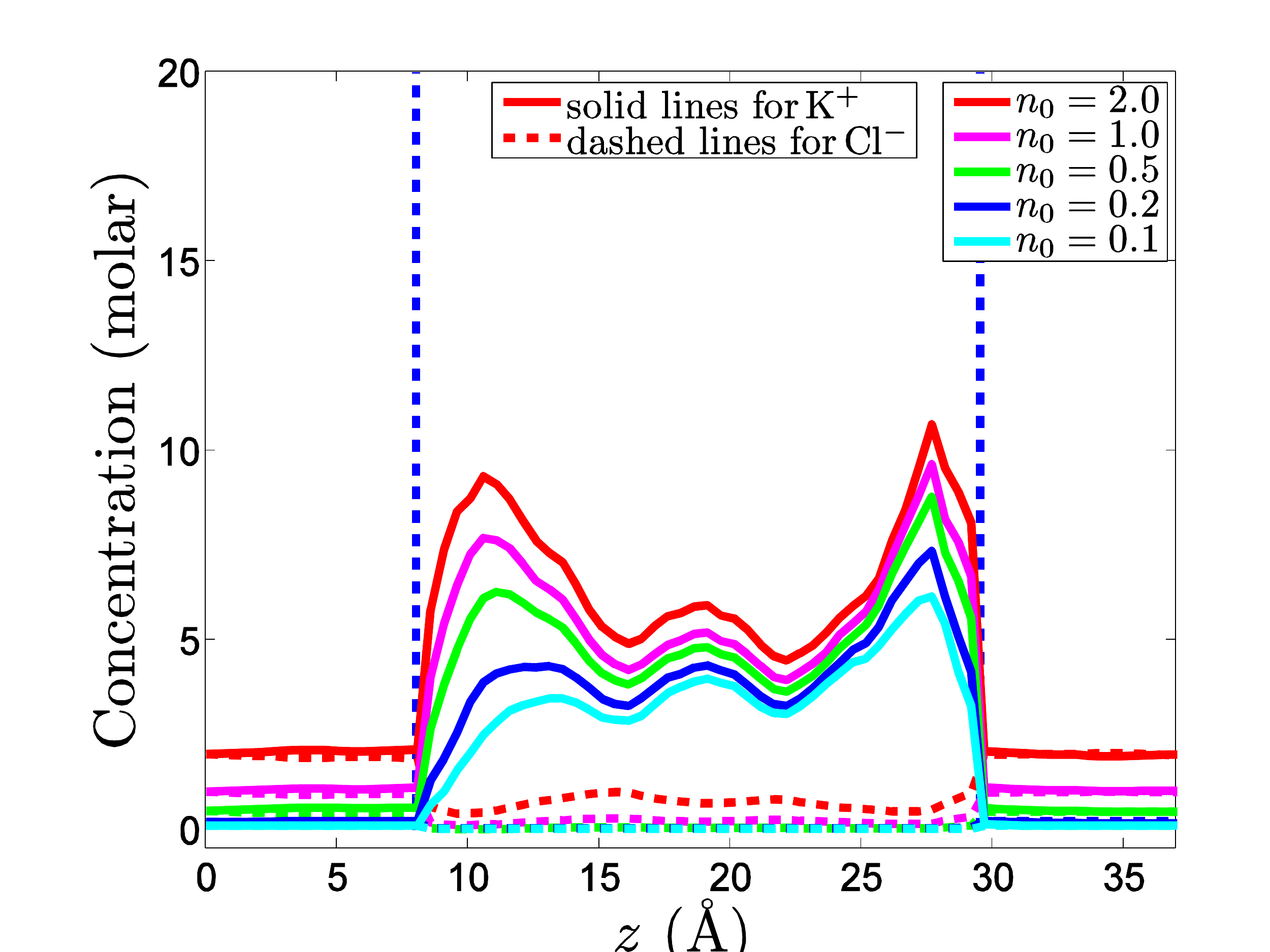}\\
                    (a)&(b)\\
                \end{tabular}
            \end{center}
             \caption{Concentration profiles in the GA channel. (a) with reference concentration $\rho_0=n_0=0.1$M and variable transmembrane voltage $\Phi_0$ (b) with $\Phi_0=50$ mV  and variable $\rho_0$. Two dashed vertical lines represent the entrance and exist of the channel.}
             \label{fig:pnp-c}
\end{figure}
	
	Concentration profiles of mobile ions in a 1:1 electrolyte, such as $\rm KCl$ or $\rm NaCl$, in the GA channel are illustrated in Fig \ref{fig:pnp-c}, where the ion concentrations of $\rm K^+$ and $\rm Cl^-$ in the GA channel are displayed,  against transmembrane voltage $\Phi_0$ and reference concentration $\rho_0=n_0$, respectively. In these figures, the concentration of $\rm K^+$ is dominant over the concentration of $\rm Cl^-$  in the channel region, so these simulations agree with the fact that GA channel only conducts cations but reject anions.
		When the reference $\rho_0=n_0$ is fixed and the transmembrane potential difference $\Phi_0$ is zero, see Fig. \ref{fig:pnp-c}(a), distribution of $\rm K^+$ concentration is somehow symmetric in the channel. While when $\Phi_0$ increases, concentration of $\rm K^+$ tends to the right end of the channel and then ion current is generated.  Fig. \ref{fig:pnp-c} shows concentration of $\rm K^+$ in channel increases as $\rho_0$ increases, while it is a constant in the bulk solvent due to the neutral electrostatics.
	\begin{figure}[ht!]
            \begin{center}
            \includegraphics[width=0.95\columnwidth]{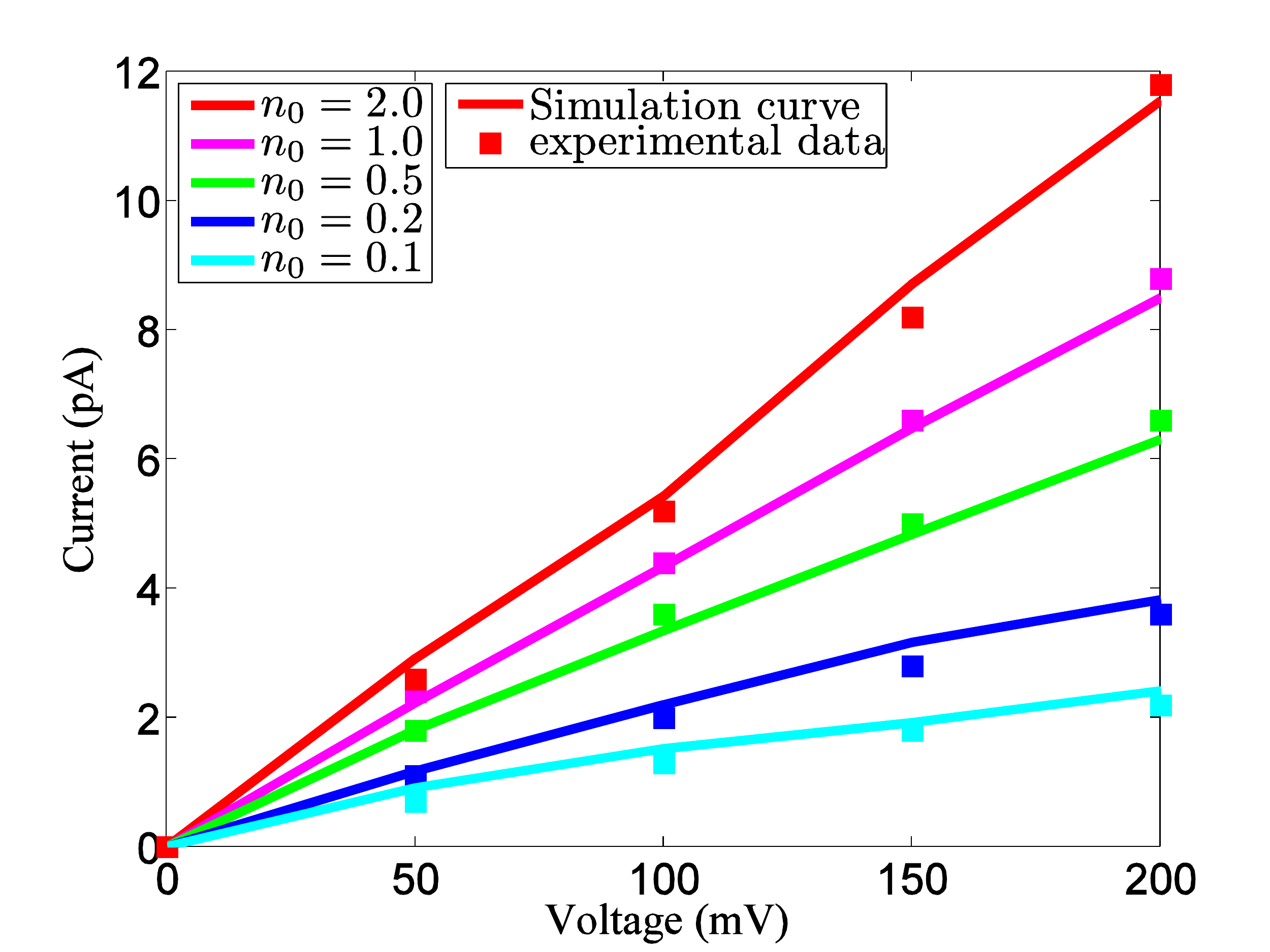}
            \end{center}
             \caption{A comparison of simulated  I-V curves and experimental data from  \cite{Cole:2002} for Gramicidin A channel.
             }\label{figiv}
	\end{figure}
	
	One way to validate the proposed model is to compare the simulated current-voltage (IV) curves with experimental results.
	In  electrophysiology, the voltage refers to the voltage across a membrane, and the current is the flow of charged ions across protein pore.  Some experimental results of I-V curves of the GA channel for KCl were reported by \cite{Cole:2002}. For the bulk diffusion coefficients of K$^+$ and Cl$^-$, the experimental data are used, i.e., $D_{\rm K} = 1.96 \times 10^{-5} {\rm cm}^2/{\rm s}$ and  $D_{\rm Cl} = 2.03 \times 10^{-5} {\rm cm}^2/{\rm s}$ for K$^+$ and Cl$^-$, respectively  (\cite{QZheng:2011a}). However, the diffusion coefficients in the channel pore are not known in general. In order to match experimental results, smaller diffusion coefficients are  usually  used in the channel region due to the restricted diffusion in most ion channels. Figure \ref{figiv} presents the  reasonable match between the IV curve simulated by the PNP  model and experimental data  from \cite{Cole:2002}, with a diffusion coefficient in the channel 25 times smaller than the bulk coefficient. 
%

\section{Generalized  PNP models}\label{sec:generalizedPNP}

\begin{figure}
		\begin{center}
			\includegraphics[width=0.95\textwidth]{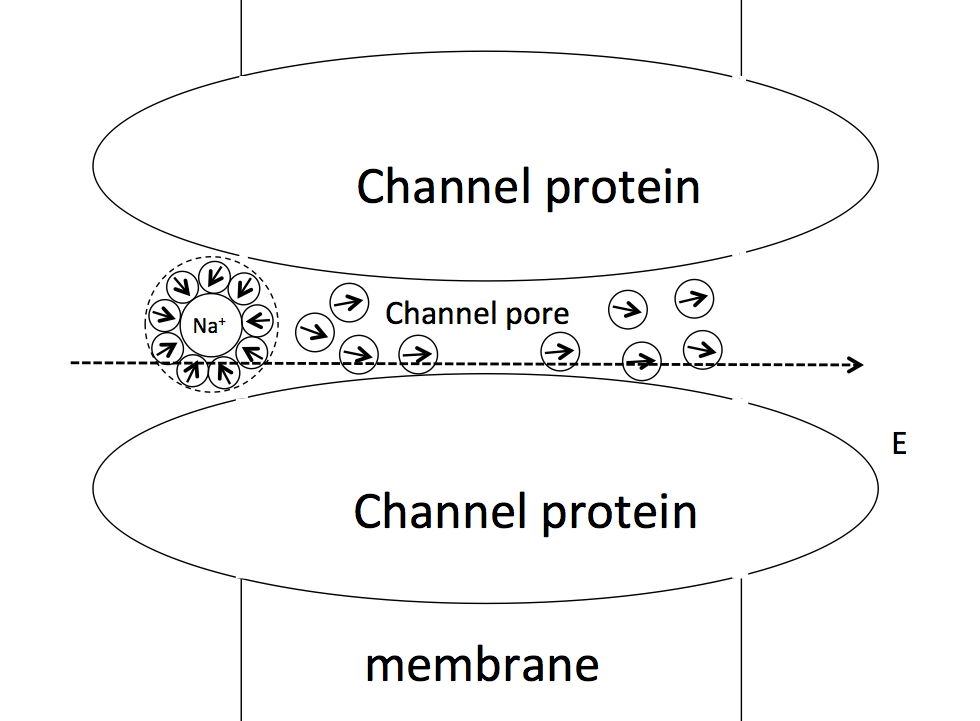}	 	
 	 	\end{center}					
 	 	\caption{(a) A schematic drawing of  steric effects of ions and ion-water interactions in channel pore. Molecular surface $\Gamma$ partitions the whole region into the solute part ($\Omega^+$) and the solvent part ($\Omega^-$). In the extremely narrow channel pore, ion-water clusters need to re-arrange their original configurations in bulk solvent in order to go through the channel region. In this situation, ion size effects and polarization of water dipoles affected in the vicinity of an ion play significant roles in ion transport dynamics. 
		 }
  		\label{fig:model}
		\end{figure}

	In the charge dynamics modeled by the traditional NP equation (\ref{eq22nernst}),  mobile ions are treated as volume-less point charges. This is a reasonable assumption for bulk or diluted solvents, but it could be problematic for crowded ionic population in a narrow channel pore. As shown in Fig. \ref{fig:model}, ionic sizes of mobile ions are comparable to geometry configuration in the narrow channel pore, and they play significant roles in interactions with water molecules and selectivity of ion channels. Many generalizations of the original PNP framework have been proposed.
Mathematical models for the finite size effects in ionic solutions were proposed by introducing  an energy term that represents the hard sphere repulsion of ions under the PNP framework, as in  \cite{Hyon:2011,Hyon:2010,lintc:2014}. The total energy then is made of the entropic energy, electrostatic potential energy, and the repulsive potential energy. As a result, modifications of the Poisson-Nernst-Planck (PNP) equations were derived, including the effects of the finite size of ions that are so important in the concentrated solutions near electrodes, active sites of enzymes, and selectivity filters of proteins. More recently, a nonlinear Poisson model, a Poisson-Nernst-Planck-Fermi  model, and an ionic concentration and size dependent dielectric permittivity Poisson-Boltzmann model were proposed to study the water molecules in solvent as heterogeneous media in the mean-field theory (\cite{LHu:2012b,liujl:2014,lub:2014}). Other work of modification of PNP theory can be found in \cite{Xuzl:2014,Qiaoyu:2014,Burger:2012, Hyon:2011, Libo:2010, Adalsteinsson:2008,Kilic2:2007}.
\subsection{Ion-size effects}
	
	A specific mathematical model for the finite size, specially, the repulsive effects of mobile ions has been introduced in \cite{Hyon:2010} for ion channels. An appropriate energy term that represents the hard sphere repulsion of ions was built into the total energy consisting of the entropic and electrostatic potential energies,  then variational approach leads to a modified NP equation for ionic species $\rho_{\alpha}$ as follows:
	\begin{eqnarray}\label{eqn:sizenp}
		\frac{\partial \rho_{\alpha}}{\partial t}&=&\nabla\cdot\left\{D_{\alpha}\left[\nabla \rho_{\alpha}+\frac{\rho_{\alpha}}{k_BT}\left( q_{\alpha}\nabla\Phi-\int\frac{12\varepsilon_{\alpha,\alpha}(a_{\alpha}+a_{\alpha})^{12}({\bf r}-{\bf r}')}{|{\bf r}-{\bf r}'|^{14}}\rho_{\alpha}({\bf r}')d{\bf r}'
		\right.
		\right.
		\right.\\\nonumber
		&-&
		\left.
		\left.
		\left.
		\sum_{\beta,\beta\ne\alpha}\int\frac{6\varepsilon_{\alpha,\beta}(a_{\alpha}+a_{\beta})^{12}({\bf r}-{\bf r}')}{|{\bf r}-{\bf r}'|^{14}}\rho_{\beta}({\bf r}')d{\bf r}'\right)\right]\right\}.
	\end{eqnarray}
	Comparing to Eq. (\ref{eq22nernst}), this model involves two integral terms with the Lennard-Jones hard sphere repulsion kernels to model the repulsing energy between ion $\rho_{\alpha}$ themselves and with all other ion $\rho_{\beta}$,
	where $\varepsilon_{\alpha,\beta}$  are  empirically chosen  energy constants for the repulsive interactions between ionic concentration $\rho_{\alpha}$ and $\rho_{\beta}$ with the radii of the two species are $a_{\alpha}$ and $a_{\beta}$, respectively.

	In another work by   \cite{Lu:2011}, the finite size effects of ions were modeled based on the Borukhov model, in which an ideal-gas-like solvent entropy term is included in the total functional, to represent the unfavorable energy of over-packing or crowding of ions in narrow channels. After variation, the new NP equation reads:
	\begin{equation}\label{eqn:sizenp2}
		\frac{\partial \rho_{\alpha}}{\partial t}=\nabla\cdot\left\{D_{\alpha}\left[\nabla \rho_{\alpha}+\frac{\rho_{\alpha}}{k_BT}\left( q_{\alpha}\nabla\Phi+\frac{k_{\alpha}\rho_{\alpha}\sum_{\beta}a^3_{\beta}\nabla\rho_{\beta}}{1-\sum_{\beta}a^3_{\beta}\rho_{\beta}}\right)\right]\right\},
	\end{equation}
	where $a_{\beta}$ is the radius of the $\beta$th ionic species and $k_{\alpha}$ is the ration between $a_{\alpha}$ and the radius of water molecules.

	In the work of \cite{Burger:2012}, a modified PNP system was established as for size effects in confined geometry of channels
	\begin{equation}\label{eqn:sizenp3}
		\frac{\partial \rho_{\alpha}}{\partial t}=\nabla\cdot\left\{D_{\alpha}\left[(1-c)\nabla \rho_{\alpha}+\rho_{\alpha}\nabla c+\frac{\rho_{\alpha}q_{\alpha}}{k_BT}(1-c)\nabla\Phi+\mu_{\alpha}\rho_{\alpha}(1-c)\nabla W_{\alpha}^0\right]\right\},
	\end{equation}
	where $\mu_i$ is the entropy variable, $c = \sum_{\alpha}\rho_{\alpha}$, and $W_i^0$ is an external potential. The nonlinear mobilities described by the model was derived based on a discrete lattice-based  hopping model with volume exclusion, combining with investigation the behavior of the system entropy in time. 
	
	In these models, the ion-ion (repulsive) interaction terms are all in terms of either concentration or gradient of concentration of mobile ions. This is the so-called density functional approach, which could be unified in a framework of density functional theory (DFT). In the differential geometry based model discussed in Section \ref{sec:GC}, the size effects of ions can be included in a unified term called generalized correlations, which include not only ion-ion interactions, but also short interactions among ions, ion-water, and ion-proteins.

\subsection{Classical  density functional theory (cDFT) based PNP (cDFT-PNP)}

Another strategy to include the above discussed size effects or other interactions beyond electrostatics is to include them uniformly by the classical density functional theory (cDFT). In this approach, the cDFT-PNP can be written as  (\cite{DMeng:2014}):
	\begin{equation}\label{eqn:cDFT}
		\frac{\partial \rho_{\alpha}}{\partial t}=\nabla\cdot\left\{D_{\alpha}\left[\nabla \rho_{\alpha}+\frac{\rho_{\alpha}}{k_BT}\left( q_{\alpha}\nabla\Phi+\nabla\mu_{\alpha}^{\rm id}({\bf r})+\nabla\mu_{\alpha}^{\rm ex}({\bf r})\right)\right]\right\}.
	\end{equation}
	Besides the electrostatics energy, the ideal chemical potential energy $\mu_{\alpha}^{\rm id}({\bf r}) $ and excess chemical potential energy $\mu_{\alpha}^{\rm ex}({\bf r}) $ are included in the ion dynamics. Both of the two terms can be expressed as functionals of ionic densities, i.e.,
	\begin{equation}\label{eqn:muid}
		\mu_{\alpha}^{\rm id}({\bf r}) = -\ln{\left[\gamma_{\alpha}\rho_{\alpha}({\bf r})/\rho_{\alpha}^0\right]},
	\end{equation}
	where $\gamma_i$ is the activity coefficient described by the extended Debye-Huckel theory. Meanwhile,
	\begin{equation}\label{eqn:muex}
		\mu_{\alpha}^{\rm ex}({\bf r}) = \frac{\delta F^{\rm ex}(\{\rho_{\alpha}({\bf r})\})}{\delta\rho_{\alpha}({\bf r})}
	\end{equation}
	and the excess chemical functional $F^{\rm ex}(\{\rho_{\alpha}({\bf r})\})$ includes hard-sphere components, short-range interactions, Coulomb interactions and electrostatic correlations, i.e.,
	\begin{equation}\label{eqn:fex}
		F^{\rm ex}(\{\rho_{\alpha}({\bf r})\}) = F^{\rm ex}_{\rm hs} + F^{\rm ex}_{\rm sh} + F^{\rm ex}_{\rm C} + F^{\rm ex}_{\rm el},
	\end{equation}
	where the expression of each term can be found in Ref.  \cite{DMeng:2014}.

\subsection{Fluid flow and chemical reactions}
When fluid flows play a crucial role in the density distribution of charge spices and electrostatic properties, the PNP equations can be coupled with the Navier-Stokes equations. 

In \cite{Eisenberg:2010,  Xusx:2014}, the Poisson-Nernst-Planck-Navier-Stokes (PNPNS) equations were derived from the energetic variational approach (EnVarA) for a 1:1 electrolyte
	\begin{equation}
		E^{\rm total}=\int_{\Omega}\left\{\frac{\rho}{2}|{\bf v}|^2+k_BT\left(n\ln{\frac{n}{n_0}}+p\ln{\frac{p}{p_0}}\right)+\frac{ze}{2\epsilon}(p-n)\int_{\Omega}G({\bf r},{\bf r}')(n-p)({\bf r}')d({\bf r}')\right\}d{\bf r}.
	\end{equation}
	Note that the energy components for electrostatics of the system is expressed by the Green's function $G({\bf r},{\bf r}')$. As results, the derived PNPNS equations are
	\begin{eqnarray}\nonumber
		&& -\epsilon\Delta\phi = ze(p-n)\\\nonumber
		&& \frac{\partial p}{\partial t} + \nabla\cdot(p{\bf r}) = \nabla\cdot\left(D_p\nabla p+\frac{ze}{k_BT}D_p p\nabla\phi\right)\\\nonumber
		&& \frac{\partial n}{\partial t} + \nabla\cdot(n{\bf r}) = \nabla\cdot\left(D_n\nabla n+\frac{ze}{k_BT}D_n n\nabla\phi\right)\\\nonumber
		&&\rho\left(\frac{\partial {\bf v}}{\partial t}+({\bf v}\cdot\nabla){\bf v}\right) = \eta\Delta{\bf v}-\nabla\Pi +(n-p)ze\nabla\phi\\\label{eqn:PNPNS}
		&&\nabla\cdot{\bf v} = 0
	\end{eqnarray}
	with the detailed formulation of $\Pi$ in Ref.  \cite{Xusx:2014}.

In another work by \cite{Wei:2009}, a total action functional was proposed as
\begin{eqnarray} \label{eqnAction}
\begin{aligned}
G^{\rm NS-PNP}_{\rm{total}}&[\Phi,\{\rho_\alpha\}] = \int\int\left\{-\frac{\epsilon_m}{2}|\nabla\Phi|^2 + \Phi \ \rho_m -\frac{\epsilon_s}{2}|\nabla\Phi|^2
+\Phi\sum_{\alpha} \rho_{\alpha}q_{\alpha} \right. \\
&\left.+\sum_\alpha\left[   \left(\mu^0_{\alpha }-\mu_{\alpha 0} \right) \rho_\alpha + k_B T  \rho_\alpha  {\rm{ln}} \ \frac{\rho_\alpha }{ \rho_{\alpha 0} } - k_B T \left(\rho_\alpha  - \rho_{\alpha 0} \right)  + \lambda_\alpha \rho_\alpha \right] \right.\\
&\left.-\left[ 
 \rho \frac{{\bf v}^2}{2} - p 
 +  \frac{\mu_f}{8}\int^{t}\left( \frac{\partial {\bf v}_i}{\partial {\bf r}_j}+\frac{\partial {\bf v}_j}{\partial {\bf r}_i}\right)^2dt' \right]\right\}d{\bf{r}}dt,
\end{aligned}
\end{eqnarray}
where $\rho =\sum_{\alpha}\rho_{\alpha}$  is the total  solvent mass density, ${\bf v}$ is the flow stream velocity, and $\mu_f$  is the viscosity of the fluid. The Einstein summation convention is used in the viscosity term.
The first few rows in Eq. (\ref{eqnAction}) have been discussed in the earlier sections. The last row in Eq. (\ref{eqnAction}) describes the Lagrangian of an incompressible viscous flow with the kinetic energy, potential energy and  viscous energy lost due to friction. Then the new NP equation derived from Eq. (\ref{eqnAction}) is
\begin{eqnarray}\label{eq22nernstFlow}
 \frac{\partial \rho_\alpha}{\partial t} + {\bf v}\cdot \nabla \rho_{\alpha} =\nabla \cdot D_{\alpha}
  \left[\nabla \rho_{\alpha}+\frac{ \rho_{\alpha}}{k_{B}T}\nabla \left(q_\alpha\Phi -  \frac{{\bf v}^2}{2} \right)\right] +\sum_j \bar{\nu}_{\alpha j}{ J}^{j},
\end{eqnarray}
where $\bar{\nu}_{\alpha j}{ J}^{j}$ is the  density production of $\alpha$ species per unit volume in the $j$th chemical reaction. Consequently, the Navier-Stokes equation results as
\begin{equation}\label{eqn:NSE44}
\rho \left( \frac{\partial {\bf v}}{\partial t} +{\bf v}\cdot \nabla {\bf v}\right)
=-\nabla p+ \nabla\cdot{\mathbb T} + {\bf F_{\rm E}},
\end{equation}
where ${\bf F_{\rm E}}=\sum_{\alpha} \rho_{\alpha}q_{\alpha}\nabla \Phi$
and  ${\mathbb T} $ is the flow  stress tensor  \cite{Wei:2009}
\begin{equation}\label{eqn:NSE2-2}
{\mathbb T}= \frac{\mu_f}{2}
\left( \frac{\partial {\bf v}_i}{\partial {\bf r}_j}+\frac{\partial {\bf v}_j}{\partial {\bf r}_i}\right)
=\frac{\mu_f}{2}\left[{\bf \nabla v} + ({\bf \nabla v})^T\right].
\end{equation}
Here the electrostatic potential is governed by the Poisson equation 
\begin{equation}\label{eqn:Poisson4}
 -\nabla\cdot\left(\epsilon \nabla\Phi \right)= \rho_m
    +\sum_{\alpha} \rho_{\alpha}q_{\alpha}.
\end{equation}
Note that  chemical reactions do not contribute to the total mass and velocity transport due to the conservation.

\subsection{Ion-water interactions}\label{subsec:GC}
	
	Another view of generalize PNP theory is to model the heterogeneous property of water molecules as a continuum. 
	Experimental observations concluded that that dielectric response of water decreases as ionic concentration increases. A possible explanation is that water molecules form a hydration shell around a solvated ion, and  when away from the ion-water cluster, the orientation of water molecules generally follow the external electrostatic field. In contrast, the motions of those dipoles in the hydration shell are greatly restricted in the vicinity of an ion. They are oriented immediately along the field line generated by the cation or anion, as shown in Fig. \ref{fig:model}, leading to an overall decrease in the dielectric response to the electrostatic field. 
	
	 Based on this observation, a PNP model involving ion-water interactions was proposed by  \cite{DuanChen:2016c} , by modeling water molecules in the solvent as a medium with dielectric function depending on the concentration of mobile ions, i.e., $\epsilon(\{\rho_{\alpha}\})$. In this work, the following total energy is considered:	

		\begin{eqnarray}\nonumber
		G[ \Phi,  \{\rho_{\alpha}\}] &=&\int_{\Omega}\left[k_BT\sum_{\alpha}\rho_{\alpha}\ln{\frac{\rho_{\alpha}}{\rho_{\alpha 0}}}- \frac{\epsilon({\bf r})}{2}|\nabla\Phi|^2+\Phi\rho_m+\Phi\sum_{\alpha}\rho_{\alpha}q_{\alpha}\right]d{\bf r}
		\\\label{eqn:FreeEnergyTotal}
		&+&
		\int_{\Omega}\sum_{\alpha}\left[k_BT(\rho_{\alpha 0}-\rho_{\alpha})+(\mu_{\alpha}^0-\mu_{\alpha 0})\rho_{\alpha}+\lambda_{\alpha}\rho_{\alpha}\right]d{\bf r},
	\end{eqnarray}
	in which the whole domain $\Omega$ has been divided into the solute domain $\Omega^+$ and the solvent domain $\Omega^-$ and
	\begin{equation}\label{eqn:dielectric}
			\epsilon({\bf r})=\left\{\begin{array}{ll}
			\epsilon_m, & {\bf r}\in\Omega^+,\\
			\\
			\epsilon(\{\rho_{\alpha}\}), & {\bf r}\in\Omega^-.
			\end{array}\right. 
		\end{equation}
	By a similar variation process, a new PNP system with ion-water interaction is obtained. 
	\begin{equation}\label{eqn:newpnp}
		\left\{\begin{array}{ll}
			\displaystyle{-\nabla\cdot\left(\epsilon(\{\rho_{\alpha}\})\nabla\Phi\right)=\rho_m+\sum_{\alpha}\rho_{\alpha}q_{\alpha}}\\
			\\
			\displaystyle{\frac{\partial  \rho_{\alpha}}{\partial t}=\nabla\cdot\left\{D_{\alpha}\left[\nabla \rho_{\alpha}+\rho_{\alpha}\nabla\left(\frac{q_{\alpha}}{k_BT}\Phi-\frac{\delta\epsilon}{\delta\rho_{\alpha}}\frac{|\nabla\Phi|^2}{2 k_BT}\right)\right]\right\}},
		\end{array}\right.
	\end{equation}
	
	Model (\ref{eqn:newpnp}) inherits the structure of  the PNP equation and introduces an extra potential energy term:
	\begin{equation}\label{eqn:ECC}
		U_{\alpha}=-\frac{\delta\epsilon}{\delta \rho_{\alpha}}\frac{|\nabla\Phi|^2}{2k_BT}.
	\end{equation}
	This energy depends on ionic species,  ionic concentration and the electrostatics, and thus it is called the ion-water interaction (IWI) energy.
	Since $\displaystyle{\frac{\delta\epsilon}{\delta \rho_{\alpha}}}<0$, the IWI energy is always positive  regardless of the charge of ions, so it is an energy barrier for all ionic species.
	It offers the ability of the model to {\em distinguish different ions of the same charges.} 
	In the conventional PNP, the overall potential energy for the  dynamics of the $\alpha$th ion is  $\Phi({\bf r})q_{\alpha}/k_BT$, i.e.,  as long as two ionic species have the same valence (e.g., $\rm Na^+$ and $\rm K^+$), they will have the identical transport dynamics, which is not realistic for some ion channels.
	In contrast, since $\delta\epsilon/\delta n_{\alpha}$ depends on specific ion types, the IWI  energy $\displaystyle{-\frac{\delta\epsilon}{\delta \rho_{\alpha}}\frac{|\nabla\Phi|^2}{2k_BT}}$ will be obviously different for ions even with the same valence.
This property can be used to study the selectivity of ion channels.
		Experimental results from   \cite{Chandra:2000} suggests that the dependence of local dielectric response $\epsilon$ of water molecules to the $\rm K^+$ is different from that of $\rm Na^+$. In other words, the value $\displaystyle{\frac{\delta\epsilon}{\delta \rho_{\rm K^+}}}$ in Eq. (\ref{eqn:ECC}) is different from $\displaystyle{\frac{\delta\epsilon}{\delta \rho_{\rm Na^+}}}$, where $\rho_{\rm K^+}$ and $\rho_{\rm Na^+}$ are local concentrations of $\rm K^+$ and $\rm Na^+$, respectively. Consequently, the overall transport potential energy, given by Eq. (\ref{eqn:newpnp}), takes different values for $\rm K^+$ and $\rm Na^+$.

\begin{figure}[ht!]
		\begin{center}
   		\begin{tabular}{cc}
  			\includegraphics[width=0.5\textwidth]{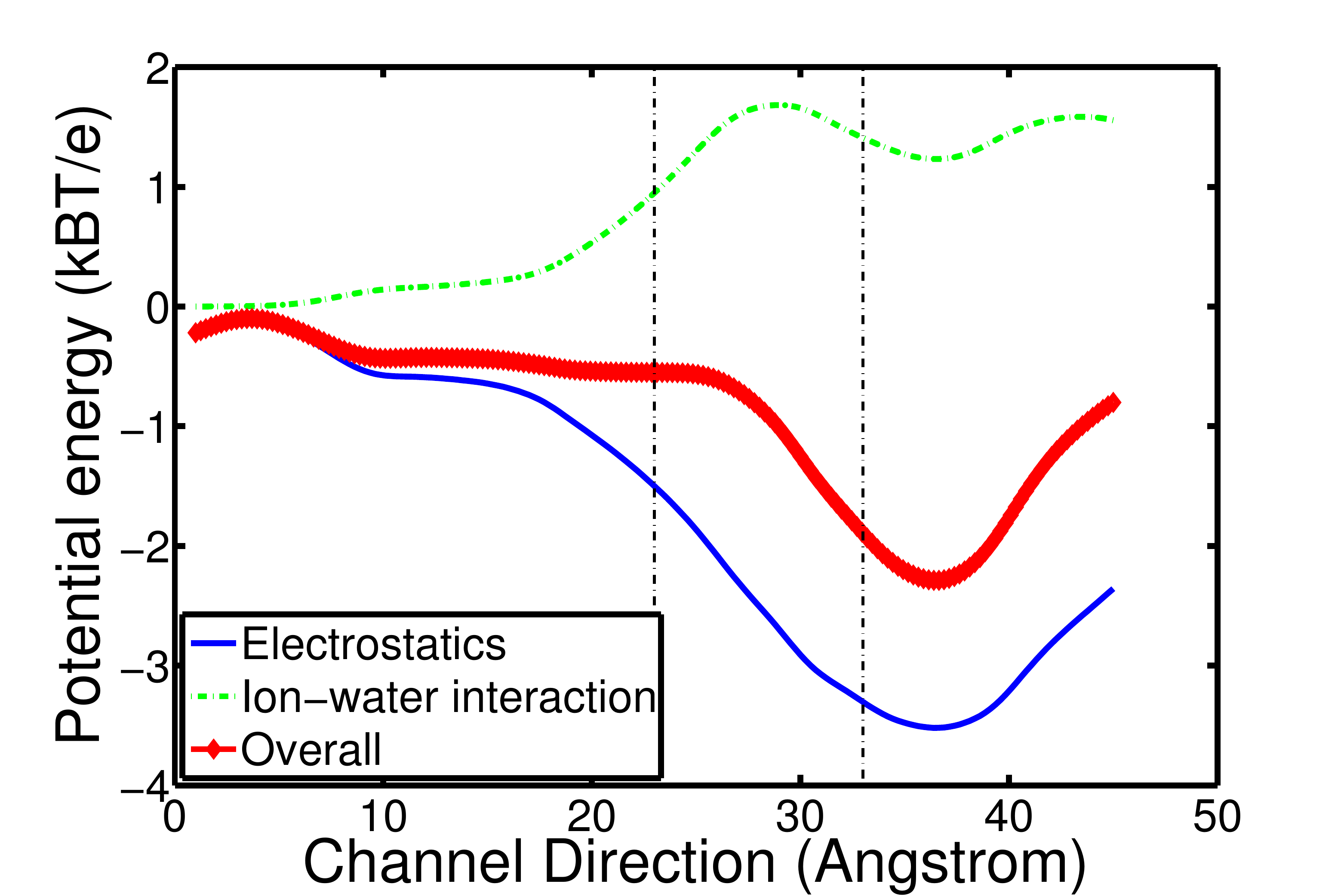}&
			 \includegraphics[width=0.5\textwidth]{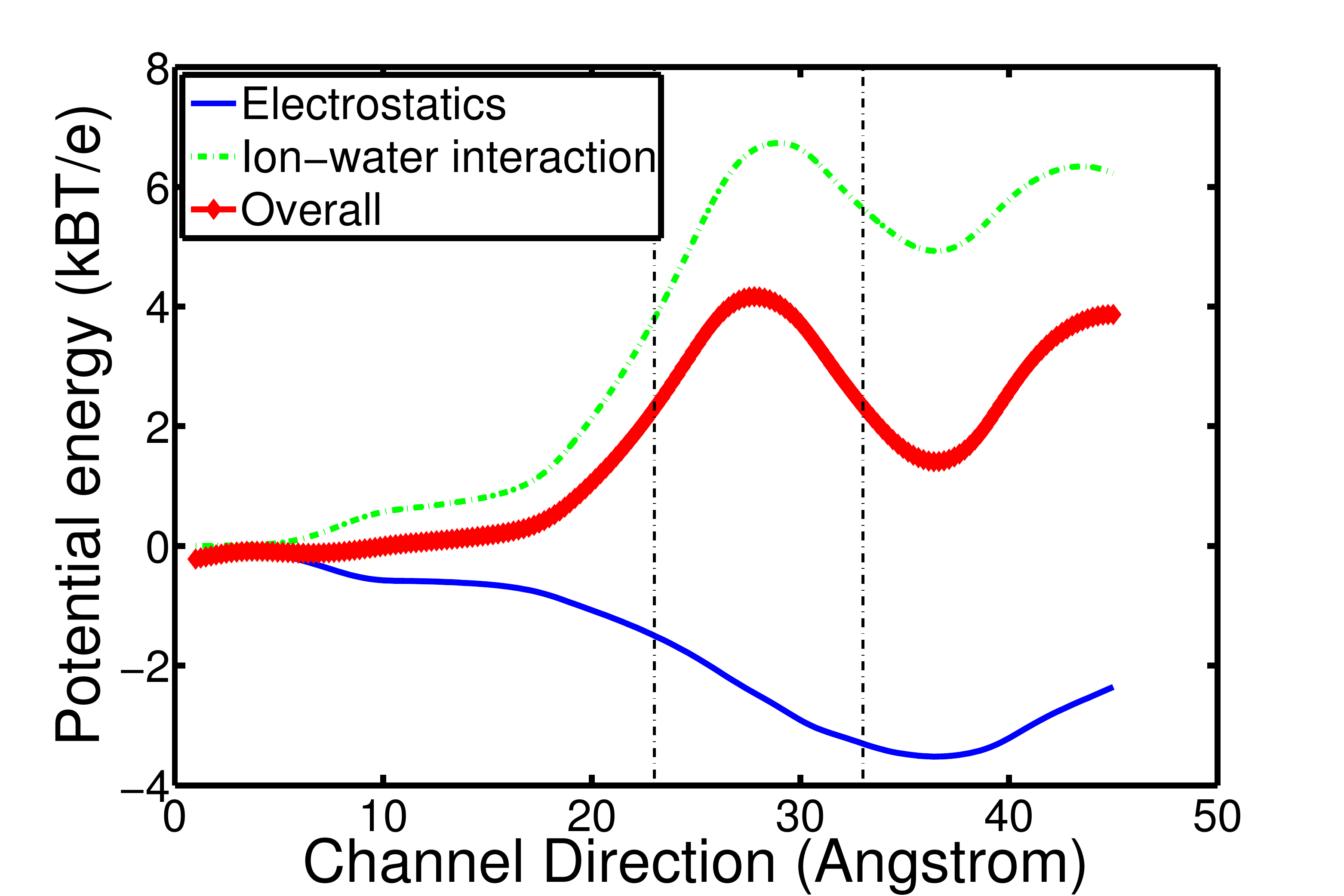}\\
			 $\rm K^+$	&	$\rm Na^+$
		  \end{tabular}
 	 	\end{center}			
 		 \caption{Profiles of potential energies of the KcsA channel simulated by the PNP-IWI model. Two dashed vertical lines separate the whole channel into three domains: the pore, cavity and filter regions. ``Overall'' means the sum of the electrostatics and ion-water interaction energies.}
  		\label{fig:kcsa-pot}
	\end{figure}

		Figures \ref{fig:kcsa-pot} (a) and (b) show the energy components  in the KcsA channel, for $\rm K^+$ and $\rm Na^+$, respectively.
	The blue curves in both figures indicate that the electrostatic potential energies are identical for $\rm Na^+$ and $\rm K^+$ (although they look differently due to the scaling). As shown  by the green curve in Fig. \ref{fig:kcsa-pot} (a), the IWI energy  of $\rm K^+$ is up to 2 $ k_BT$  and the overall, effective transport energy  (red curve) is alleviated through the channel, but is still negative. On the other hand, $\rm Na^+$ ions experience the IWI energy as high as 8 $k_BT$, so the overall potential energy in the KcsA channel becomes an energy barrier. It is worth to pointing out that the dielectric functions and values of the parameter $\beta$ for $\rm K^+$ and $\rm Na^+$ are taken from experiment results. Therefore, the proposed formulation is able to model the selectivity of the KcsA channel, although the reality is much more complicated.
	
\begin{figure}[ht!]
		\begin{center}
  			\includegraphics[width=0.95\textwidth]{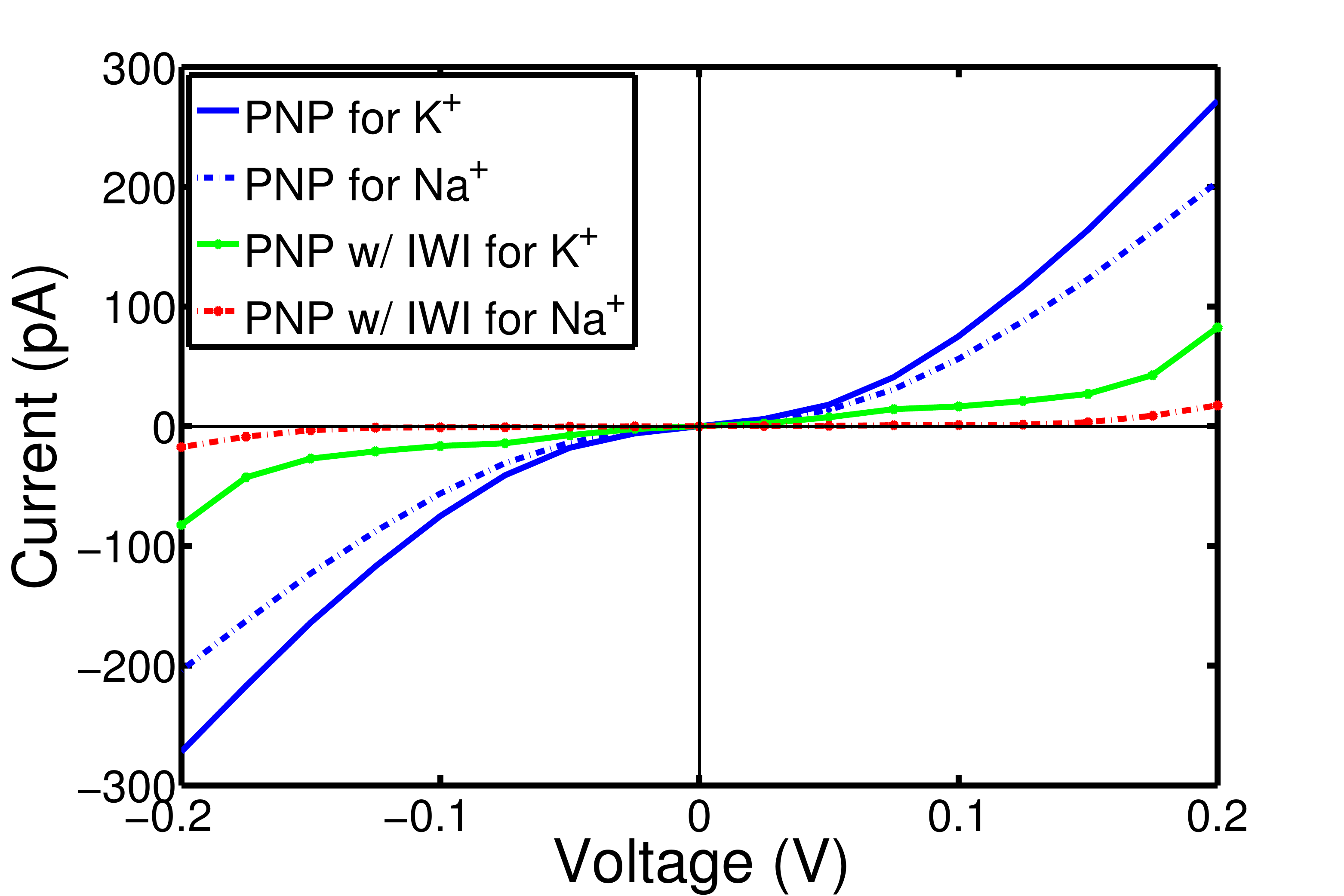}
 	 	\end{center}			
 		 \caption{Simulated current-voltage relations of $\rm K^+$ and $\rm Na^+$ in the KcsA channel.}
  		\label{fig:kcsa-iv}
	\end{figure}

	In Fig. \ref{fig:kcsa-iv}  the current-voltage relations of KcsA channel are compared for the fundamental PNP model and the PNP with IWI model. As shown by blue curves, the PNP model predicts similar magnitudes of ionic currents for $\rm K^+$ and $\rm Na^+$; the only difference is from the diffusion coefficients of two ion species ($\rm K^+$: $1.96\times 10^{-9}m^2/s$ and $\rm Na^+$:$1.33\times 10^{-9}m^2/s$  \cite{Kuyucak:2001}). This is against the biological observation that the conductance of $\rm K^+$ in the KcsA channel is dominant over $\rm Na^+$. Secondly,   it is widely believed that PNP model generally overestimates ionic current and a phenomenologically  tuned diffusion coefficient is used to match experimental results. The green and red curves are the current-voltage generated by the PNP-IWI model for $\rm K^+$ and $\rm Na^+$, respectively. From these curves, one can see that the simulated ionic current of $\rm K^+$ is significantly higher than  that of $\rm Na^+$. Additionally, diffusion coefficients do not need to be reduced in the PNP-IWI model.

\subsection{Poisson-Boltzmann-Nernst-Planck model}\label{sec:PBNP}
	
		The PNP model provides good descriptions of each ionic species in a non-equilibrium system for charge transport phenomena. However, the computational cost will be extremely high when there is a large number of ionic types in the system because each charge species is governed by one 3D Nernst-Planck equation that needs to be numerically solved. To address this issue, a Poisson-Boltzmann Nernst-Planck (PBNP) model was introduced by  \cite{QZheng:2011b}, in which only  concentrations of the target ions (ions of interests) are modeled by the Nernst-Planck equation while those of other ions are described by the Boltzmann distribution. This approach is especially reasonable for charge transport modeling: an ion channel usually has selectivity and it only conducts some specific ion species. For the rest of ions, they can be considered in a quasi-equilibrium state in the bulk solvent. 
		 The validity and usefulness of the PBNP formulation were confirmed by independent researchers such as   \cite{Kiselev:2011}.  


Assume  the total number of ion species in the system is $N_c$, among which $\rho_{\alpha} ~(\alpha= 1,\cdots, N_{\rm NP})$ are denoted as the densities of the target charge species, or ions of interests, thus $N_{\rm NP}$ is the total number of charge species treated by using the non-equilibrium Nernst-Planck (NP) equation.  On the other hand, the densities of the remaining charge species in the system are $\rho_{\beta} ~ (\beta=N_{\rm NP}+1,\cdots,N_c)$  and  $N_{\rm BD}=N_c-N_{\rm NP}$ is the total number of the remaining charge species which are represented by the equilibrium Boltzmann distribution.  
Based on this consideration, the total free energy functional can be expressed by
\begin{eqnarray}  \label{eq34energy}
\begin{aligned}
G^{\rm PBNP}_{\rm{total}}[\Phi,\{\rho_\alpha\}]&= \int \left\{-\frac{\epsilon_m}{2}|\nabla\Phi|^2 + \Phi \ \rho_m-\frac{\epsilon_s}{2}|\nabla\Phi|^2+\Phi \sum_{\alpha=1}^{N_{\rm NP}}  \rho_{\alpha}q_{\alpha}
-\sum_{\beta=N_{\rm NP}+1}^{N_c} k_B T \rho_{\beta 0} \left(e^{-\frac{q_{\beta }\Phi-\mu_{\beta0}}{k_B T }} -1 \right)
\right.  \\
& \left. +\sum_{\alpha=1}^{N_{\rm NP}}
\left[  \left(\mu^0_{\alpha } - \mu_{\alpha 0} \right)\rho_\alpha + k_B T  \rho_\alpha  {\rm{ln}} \ \frac{\rho_\alpha }{ \rho_{\alpha 0} } - k_B T \left(\rho_\alpha  - \rho_{\alpha 0} \right)  + \lambda_\alpha \rho_\alpha \right]
\right\}
d{\bf{r}}.
\end{aligned}
\end{eqnarray}
Equation (\ref{eq34energy}) includes the same energy components as Eq. (\ref{eq17tot}) does, but note two different treatments for charged ion species are taken for the charge source terms in the polar solvation free energy functional.  


It is then a standard procedure to derive the  PBNP equations from the total energy functional (\ref{eq34energy}).
\begin{equation}\label{eqn:newpnp}
		\left\{\begin{array}{ll}
			\displaystyle{ - \nabla\cdot\left(\epsilon \nabla\Phi \right)= \rho_m
    +\left(\sum_{\alpha=1}^{N_{\rm NP}} q_{\alpha} \rho_{\alpha} + \sum_{\beta=N_{\rm NP}+1}^{N_c} q_{\beta}\rho_{\beta 0} e^{-\frac{q_{\beta }\Phi  -\mu_{\beta0}}{k_B T }}, 
    \right).
}\\
			\\
			\displaystyle{\frac{\partial \rho_\alpha}{\partial t}=\nabla \cdot \left[D_{\alpha}
  \left(\nabla \rho_{\alpha}+\frac{ \rho_{\alpha}q_\alpha}{k_{B}T}\nabla \Phi \right)\right]},
		\end{array}\right.
	\end{equation}

In practical applications, the Nernst-Planck equation is only needed for the ions of interests $\rho_{\alpha}$ and usually $N_{\rm NP}<<N_c$.
This treatment can significantly reduce computational costs in simulations for a given level of modeling accuracy.
	For an electrolyte that contains several ionic species, it is strategically useful to focus on the ion of major interests and then use the PBNP equations, instead of the full PNP equations, in order to reduce model complexity and computational costs. The ability of quasi-equilibrium PBNP model   to recover the full predictions of the non-equilibrium PNP model  was tested in  \cite{Wei:2012}. 
		
\begin{figure}[ht!]
            \begin{center}
                \begin{tabular}{cc}
                        \includegraphics[width=0.5\columnwidth]{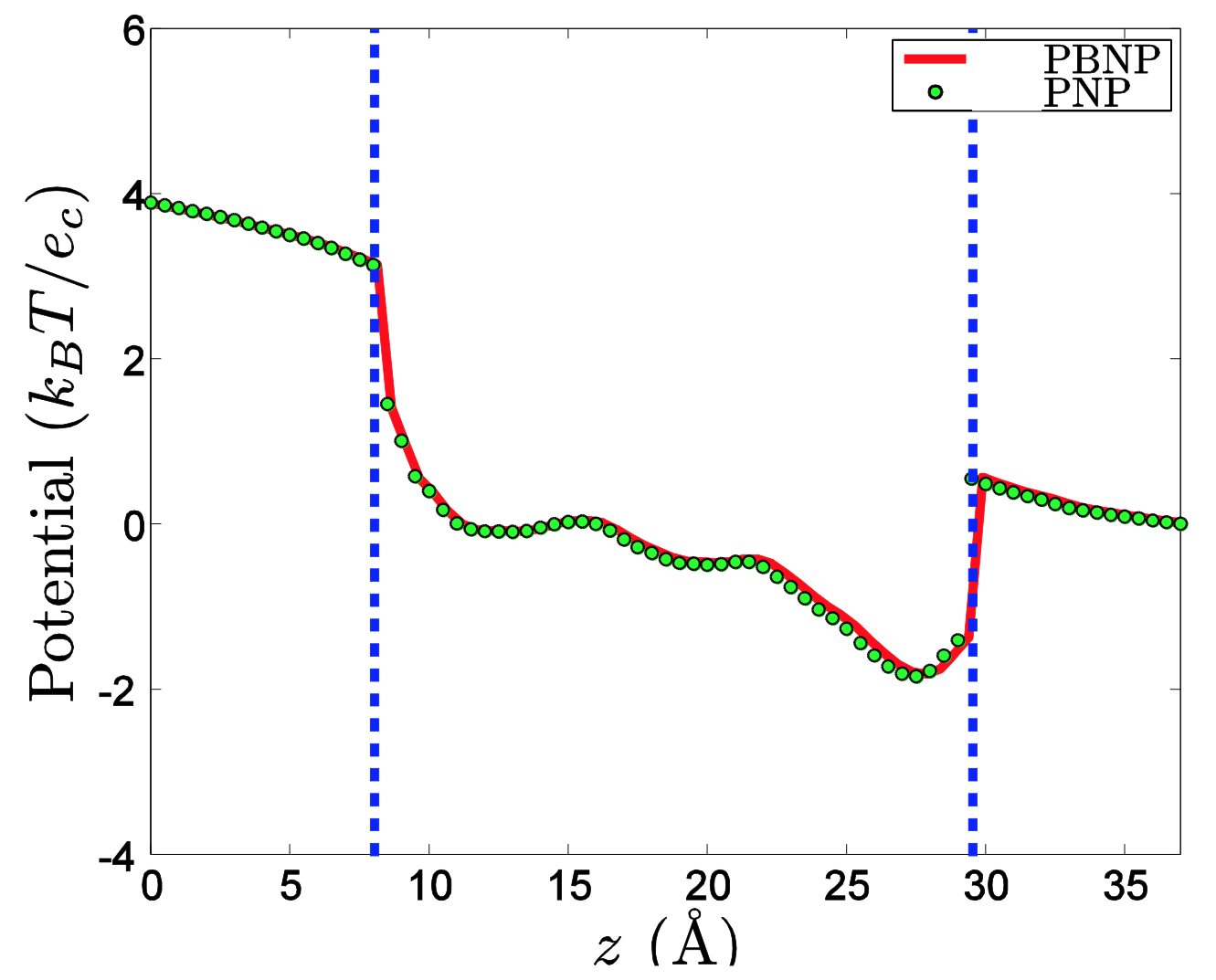}&
                    \includegraphics[width=0.5\columnwidth]{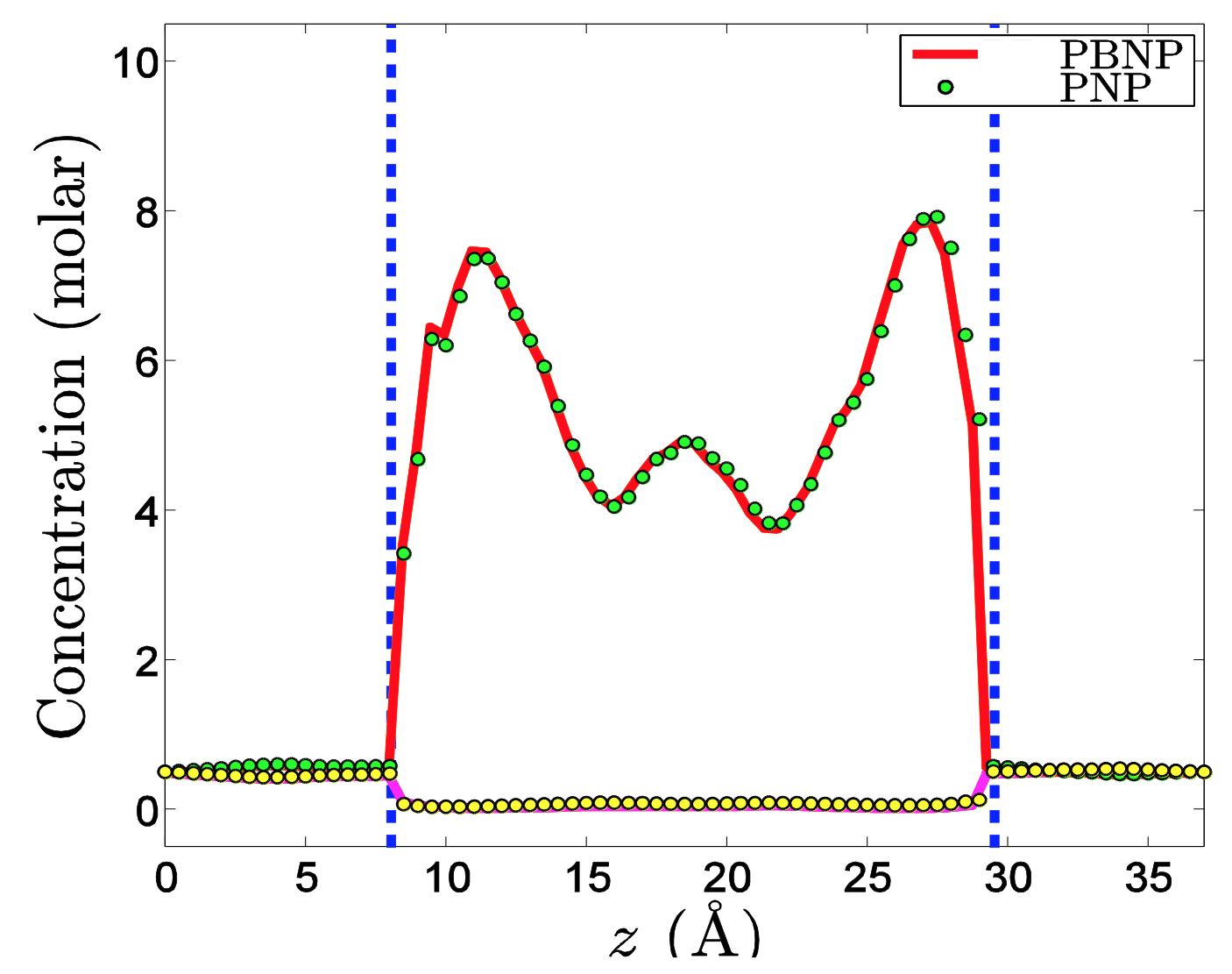}\\
                    (a)&(b)\\
                \end{tabular}
            \end{center}
             \caption{Comparison of cross sections of electrostatic  potential and concentration profiles from the PNP and PBNP models.  Transmembrane voltage $\Phi_0=100$$\rm{mV}$, reference concentration $n_0=0.5$ M.
             (a) Electrostatic potential profiles;
             (b) Concentration profiles. }\label{figv1}
\end{figure}

		Figure \ref{figv1} provides the comparison of the cross sections of electrostatic  potential and concentration profiles obtained from  PNP and  PBNP models (\cite{QZheng:2011b}). The external voltage is set to $\Phi_0=100$$\rm{mV}$ and the salt (KCl) concentration is $n_0=0.5$M. Concentration of Cl$^-$, $n_{{\rm Cl}^-}({\bf r})$ is  represented   by using the Boltzmann distribution, while solve the Nernst-Planck equation for K$^+$ density $n_{{\rm K}^+}({\bf r})$.  In figures, solid curves are simulation results from the PBNP equation, while the dots are from the PNP equations. Electrostatic potential computed by the reduced LB-PBNP model agree quite well with the full  PNP model. For the density profile, reduced PBNP model also does an excellent job in the channel region, which is the region of main interest.
\section{Computational algorithms and implementation}\label{sec:algorithm}

	In order to understand the realistic chemical, physical, and biological properties in ion transport processes and to predict reliable results, many algorithms and computational tools have been developed to obtain highly accurate and efficient numerical solutions of various proposed models. 
	

	
\subsection{Finite difference based methods}

	The PNP and related system can be discretized by finite difference method (FDM). In \cite{QZheng:2011a,DuanChen:2011a}, a set of FDM based algorithms were developed for the second-order convergence solutions of the PNP equations with 3D {\em realistic and complicated} solvent excluded surfaces, in which the matched interface and boundary method (MIB) developed in \cite{Yu:2007c,Yu:2007a,Zhao:2004,Zhou:2006c,Zhou:2006d, Zhou:2008b, KLXia:2011, KLXia:2012a, KLXia:2014e, KLXia:2014f} was used to handle the discontinuous property of dielectric constants on solvent and solute domain, and Dirichlet to Neumann mapping technique  (\cite{Geng:2007a}) was applied to rigorously treat Dirac delta singularities of fixed charges the protein channels.   
This PNP algorithm was used to construct 	a molecular level  prototype for  mechanoelectrical  transducers in  mammalian   hair cells  (\cite{JKPark:2013a}).
Finally, 	many   PB solvers, as those in \cite{Bertonati:2007,Rocchia:2001,Baker:2001b,CHARMM-PB,LuoRayPBSA:2012}, can be incorporated to PNP systems.

	Finite-difference methods for solving 1D and 2D time-dependent PNP equations was developed  by \cite{lixf:2014, LiuHL:2014}, with second-order accurate solutions in both space and time. These work focus on conservation of total ions, correct rates of energy dissipation, and positivity of the ion concentrations. A set of sufficient conditions on the step sizes of the numerical method were discussed to assure positivity of the ion concentrations and it demonstrated that  the conservation property is critical in obtaining correct numerical solutions over long time scales (\cite{lixf:2014}). Relatively simple and easy-to-implement conservative schemes were established to preserve equilibrium solutions, and they were proved to satisfy the total exact concentrations, preserving positivity of the chemical concentrations under a mild CFL condition, and the free energy dissipation law at the semi-discrete level  (\cite{LiuHL:2014}).

	Efficient numerical algorithms for solving 3D steady-state PNP equations with excess chemical potentials described by the classical density functional theory (cDFT)  (\cite{DMeng:2014}). In these algorithms, the NP equations were transformed into Laplace equations through the Slotboom transformation. The algebraic multigrid method was applied to efficiently solve the system and excess chemical potentials was calculated through fast Fourier transforms with computational complexity of $O(N \log N)$, where $N$ is the number of grid points. 

Pseudo-time-coupled nonlinear models was proposed for biomolecular surface representation	and solvation analysis  by \cite{SZhao:2011}. Recently, this type of approaches have been further extended to operator splitting  alternating direction implicit (ADI) schemes for pseudo-time coupled nonlinear solvation simulations  (\cite{SZhao:2014a}). More recently, a fast {ADI} algorithm has been developed for geometric flow equations in biomolecular surface generations  in \cite{WFTian:2014}.

\subsection{Finite element based methods}
	Finite element method (FEM) is also developed to solve the PNP equation  (\cite{Lu:2010, Lu:2011,BTu:2013, Chaudhry:2014,Sunyz:2016,Metti:2016}). An FEM based solver was developed for a modified form of the PNP equations that includes steric effects of mobile ions by \cite{Chaudhry:2014}. The algorithm in this work combines  the Newton's method to the nonlinear Galerkin form of the equations, which are augmented with stabilization terms to appropriately handle the drift-diffusion processes. Periodic boundary conditions of the PNP equations were used to conserve the number of ions in the solution domain and to make comparison with particle-based simulations possible.
	
A stable regularization scheme was applied to remove the singular component of the electrostatic potential induced by the permanent charges inside biomolecules, and then regular, well-posed weak form of PNP equations were formulated by \cite{Lu:2010,Lu:2011},. For the steady problems, an inexact-Newton method was used to solve the coupled nonlinear elliptic equations, and  for time integration for the unsteady electrodiffusion, the Adams-Bashforth-Crank-Nicolson method was devised. These computational algorithms were also generalized to a size-modified Poisson-Nernst-Planck (SMPNP) model that is able to treat nonuniform particle sizes by the Borukhov model.
	
	A method of lines approach was proposed for  the FEM discretization to for approximately solve the PNP equations in   \cite{Metti:2016}. This discretization scheme assures positivity of the numerical solutions, which have physical meaning of particle density. A discrete energy estimate was also established and extended to the FE solutions of an electrokinetic model, which couples the PNP system with the incompressible Navier-Stokes equations. A parallel FE simulator for ion transport was developed in \cite{BTu:2013} and error analysis of the FEM method for the PNP equations is available in  \cite{Jerome:1990,Sunyz:2016}.

\subsection{Other computational methods available for ion channels}	
 
 	There are many other intelligent computational algorithms, some of which are hybrid methods, and some are currently developed  for  the equilibrium PB equation but could be extended to study non-equilibrium PNP-like equations in the future.
	
	{\em Hybrid models:}  \cite{Caiw:2013, Caiw:2016} developed  an image-charge solvation method (ICSM)  combined with molecular dynamics simulations  to investigated the selectivity of the KcsA channel. In this hybrid model, all particles including the channel protein, water molecules and mobile ions were described at the atomic level with molecular dynamics in a small neighborhood around the channel, while the reaction field effect of the continuum approximation of the background was computed by the image method. The ICSM is able to demonstrate the function of the selectivity filter of the KcsA channel when potassium and sodium ions are considered. In  \cite{Jung:2009}, a PNP with explicit resident ion, or ERIPNP model, was developed to study biting sites of $\rm K^+$ ions in the KcsA channel. In this algorithms, the continuum PNP equations are accompanied by explicitly described individual ions with finite size in the selectivity filter of the channel. The ERINP model reproduced the experimental results with a realistic set of parameters and also reduced CPU costs.

	{\em Boundary integral method:}. A boundary integral equation program was provided  for calculation of electrostatics in the Poisson Poisson-Boltzmann modeling of an ion channel in layered dielectric/electrolyte media  in \cite{Caiw:2013, Zinser:2016}. A layered media Green's function was used   in order to accurately model the inhomogeneous background, including different physical properties and extra/intra cellular environments, cell membrane, and the cylindrical shape of ion channels. A series of parallel, treecode, or GPU-accelerated boundary integral equation methods were developed to calculate electrostatics of solvated biomolecules  by  \cite{Geng2013:1,Geng2013:2,Geng2013:3}.
	
	{\em Nonlocal algorithms:} A nonlocal dielectric model and the associated computational package was developed for protein in ionic solvent, taking into account the polarization correlation among water molecules in  \cite{Xie:2007,xied:2013}. Using solution splitting and reformulation techniques, the solution of the nonlocal dielectric model was shown to be uniquely obtained  from solving two well-defined PDE systems and one Poisson-like boundary value problem. Additionally, a nonlocal linearized PB equation with uniform ionic size effect was also proposed and numerically tested on three protein molecules.
	
	{\em Stochastic methods:} Stochastic walk-on-spheres (WOS) algorithms for solving the linearized Poisson-Boltzmann equation (LPBE) provide several attractive features not available in traditional deterministic solvers (\cite{Mascagni:2007}): Gaussian error bars can be computed easily, the algorithm is readily parallelized and requires minimal memory and multiple solvent environments can be accounted for by reweighting trajectories. Numerical optimizations that can make the computational time of Monte Carlo LPBE solvers competitive with deterministic methods was introduced   by \cite{Mascagni:2013}. In the optimization techniques, each atom's contribution to the variance of the electrostatic solvation free energy was assured to be the same, and the bias-generating parameters in the algorithm were   optimized, and an epsilon-approximate rather than exact nearest-neighbor search was utilized when determining the size of the next step in the Brownian motion.

\subsection{Iterative schemes for coupled systems}

Regardless computational methods, governing equations in all models are coupled and need to be solved iteratively. To illustrate the iterative process, consider a coupled generalized PNP model as
 \begin{eqnarray}\label{eqn:sys-lb}
    &&  \frac{\partial S^{k+1}}{\partial t}=|\nabla S^{k+1}|\left[\nabla\cdot\left(\gamma\frac{\nabla S^{k+1}}{|\nabla S^{k+1}|}\right)
   + V_{\rm LB}(\Phi^k, \{\rho^k_{\alpha}\})\right],\\\label{eqn:sys-ps}
    &&-\nabla\cdot\left(\epsilon(S^k) \nabla\Phi^k \right)= S^k\rho_m
    +(1-S^k)\sum_{\alpha} \rho^k_{\alpha}q_{\alpha}, \\\label{eqn:sys-np}
    &&\nabla \cdot \left[D_{\alpha}
  \left(\nabla \rho^k_{\alpha}+\frac{ \rho^k_{\alpha}}{k_{B}T}\nabla (q_\alpha\Phi^k +U_{\alpha})\right)\right]=0,
\end{eqnarray}
where the function $S({\bf r})$ is a characteristic function identifying the solvent or solute domain. Derivation of the governing equation of $S({\bf r})$, or Eq. (\ref{eqn:sys-lb}) is discussed in detail in Section \ref{sec:DG}.  In this case, solving Eq. (\ref{eqn:sys-lb}) is considered as the outer loop, in which the function $S^{k+1}$ in the $(k+1)$th step is solved from the electrostatics $\Phi^k$ and $\rho^k_{\alpha}$ in the $k$th step. On the other hand, solving for $\Phi^k$ and $\rho^k_{\alpha}$ with an available $S^k$ is also an iterative process and it is called the inner loop. Namely, start with a certain definition of molecular surface $S^0$, $\Phi^0$ and $\rho^0_{\alpha}$ are solved self-consistently with this chosen $S^0$ till convergent.  In this case, $\Phi^{0,0}$ is taken as the solution of the corresponding Poisson-Boltzmann equation, and it is used in Eq. (\ref{eqn:sys-np}) to solve for $\rho^{0,1}_{\alpha}$. Then $\rho^{0,1}$ is substituted in Eq. (\ref{eqn:sys-ps}) to solve for $\Phi^{0,1}$. After $m$th step $\Phi^{0,m}$ and $\rho^{0,m}_{\alpha}$ are convergent, they are renamed as $\Phi^0$ and $\rho^0_{\alpha}$ and used in the outer loop Eq.(\ref{eqn:sys-lb}) to solve $S^1$.

In the $m$th inner loop for computing $\Phi^{k,m}$ and $\rho^{k,m}_{\alpha}$, the successive over relaxation scheme is utilized  ( \cite{ZhanChen:2010a})
\begin{eqnarray}\label{eq46relax}
\begin{aligned}
\Phi^{k, m}&=&\zeta_1 \Phi^{k,m} + (1-\zeta_1)\Phi^{k, m-1}\\
\rho_{\alpha}^{k, m}&=&\zeta_2 \rho_{\alpha}^{k,m} + (1-\zeta_2)\rho_{\alpha}^{k, m-1},
\end{aligned}
\end{eqnarray}
where $0\le\zeta_1\le 1$ and $0\le\zeta_2\le1$ are relaxation factors.  Larger values of $\zeta_1$ and $\zeta_2$ will lead to slower convergence, while  smaller values may cause instability. Alternatively, the Gummel iteration  proposed by \cite{Gummel:1964} can also be used to handle this type of problems.

The overall self-consistent process, including inner and outer iterations are summarized as follows.
\begin{itemize}
\item[]
Step 1: Initial atomic position and partial charge generation.
The initial atomic positions of a protein are taken from the Protein Data Bank (PDB) (www.pdb.org), and  the partial charge prescription is obtained by the software PDB2PQR   \cite{Dolinsky:2004,Dolinsky:2007}, which provides ${\bf{r}}_j$ and  $Q_j$  values in the formulation.
\item[]
Step 2: Given initial guesses of $\Phi$ and $\rho_\alpha$, the surface function $S$ is obtained by the initial value problem Eq. (\ref{eq25surf}).
After the surface function $S$ is determined, an isosurface is extracted for the interface $\Gamma$.
\item[]
Step 3:  Based on the interface $\Gamma$,  normal direction ${\mathbf n}$ is computed by $\frac{\nabla S}{|\nabla S|}$ on the isosurface; the coupled Eqs. (\ref{eqn:sys-ps}) and (\ref{eqn:sys-np}) are solved iteratively by above mentioned schemes.
\item[]
Step 4:  Go to Steps 2 and 3 for updating $S$, $\Phi$ and $\rho_\alpha$ until a convergence is reached based on a given tolerance. Noticed that in the $l$th outer loop for updating $S$, we use $S^{l+1}=\lambda_3 S^{l} + (1-\lambda_3)S^{l+1}$. In each outer loop, the total free energy functional is evaluated for checking the convergence criteria.
\end{itemize}

\section{Mathematical analysis}\label{sec:analysis}

The mathematical analysis established for the PNP and related models in past decades are majorly in two aspects: one is the asymptotic behavior of the PNP model and the other is the multiple solutions of modified PNP equations.

The PNP system was studied as a singularly perturbed system, with the assumption that the Debye length is small relative to the diameter of the narrow ion channel, by  \cite{LiuWS:2005,Abaid:2008,Eisenberg:2006, Singer:2008}.  Explicit derivation of higher order terms in the asymptotic expansions was obtained from special structures of the zeroth order inner and outer systems. Various current-voltage relations of ion channels were described in the case of zero permanent charges in a channel protein with electro-neutrality condition enforced at the ends of the channel   (\cite{Abaid:2008,Zhangmj:2015}).  Ion channels involving two types of ions with three regions of piecewise constant permanent charge were studied by geometric singular perturbation theory, which gives rise to the existence and (local) uniqueness of the solution of the singular boundary value problem near each singular orbit  (\cite{Eisenberg:2006}). By this technique, multiple solutions of the system were discovered and they might explain a variety of multiple valued phenomena in biological channels, such as gating or  some kinds of active transport.  \cite{LiuWS:2005} studied the global behavior in terms of limiting fast and slow systems, in which three different time scales were indicated in the singularly perturbed PNP system.

Solutions of modified PNP system may explain more physical observation. The ability of PNP system is analyzed to study gating mechanism of ion channels by   \cite{lintc:2015},.  Discovered by experimental measurements,  single protein channels produces unstable currents: nearly zero or a definite level of currents. One reason may be the spontaneous stochastic gating process. Existence of multiple solutions of steady state PNP-steric equations were studied to check whether it can describe this two levels of current. Indeed,  two steady state solutions of PNP-steric equations were proved for three types of ion species (two types of cations and one type of anion) and four types of ion species (two types of cations and their counter-ions) with specific assumptions on permanent charges in channel proteins. In \cite{Zhangmj:2016}, a quasi-one-dimensional steady-state PNP model modified with size effect was studied as a singularly perturbed differential system, with fixed boundary ion concentrations and electric potentials. The existence of solutions to the boundary value problem for small ion sizes was investigated with the ion sizes as small parameters. This analysis provided dependencies of  current-voltage relations on boundary concentrations, diffusion coefficients and ion sizes.

\section{Other mathematical models for  ion channel transport   }\label{sec:others}

\subsection{Poisson-Boltzmann-Kohn-Sham model for proton transport}\label{sec:PBKS}

	Proton transport is one special type of charge transport through membranes and plays a  critical role in many biochemical processes  (\cite{DuanChen:2011c, DuanChen:2011d}). For example, generation and conduction of large proton concentration gradients are required in energy transduction in a bioenergetic system: chemical energy is stored as proton gradient that drives the ATP generation in mitochondria of animal cells. While for plants, light energy is transducted into a proton gradient to create ATP in chloroplasts    (\cite{Decoursey:2003}). 
	Studies of proton transport  are also important for public health: for example, the M2 proton ($\rm H^+$) channel in influenza A virus. Conducting protons into the virion core and thus acidifying the virus interior, the M2 proton channel leads viral ribo nucleo protein (RNP) complexes release and start viral replication  (\cite{Schnell:2008}). Also, it has been intensively studied that proton channels are frequent targets in the research of new drugs for human diseases such as cancers  (\cite{Harguindey:2009}).
	
	However,   the mechanism  proton transport is  significantly different from that of regular ions such as sodium or potassium. The reason is that proton has the lightest mass among all ions and its effective radius at least $10^5$ smaller  because the $\rm H^+$ has no electron  (\cite{Decoursey:2003}). The light mass and tiny size greatly facilitate proton transfer reaction and electrostatic interactions with surrounding molecules  (\cite{Mitchell:1976}). Due to these unique physical properties, the mobility of protons in bulk solution is about fivefold higher than that of other cations   (\cite{Bernal:1933}). 
	 Main mechanism of proton transport is not fully understood yet   (\cite{Till:2008}), but major studies indicated that transport of protons is characterized as a succession of hops in the hydrogen-bond network and is described by the Grotthuss theory   in \cite{Nagle:1978}.

Proton transport needs to be treated by quantum mechanical formulations, according to \cite{Nagle:1978,Roux:1996}. Many algorithms, such as multistate empirical valence bond (MS-EVB) approach in \cite{Schmitt:1999},  were developed by Schmitt and Voth,  to compute dynamics of protons in bulk phase water with an emphasis on a quantum dynamical treatment. Additionally, using Feynman path integral dynamical simulations in  \cite{Pomes1:1996, Roux:1996, Roux2:2002}, Roux and his coworkers investigated  single file of water molecules in the Gramicidin A channel, which functions as a proton wire. Although not a governing factor,  the nuclear quantum effect has a significant  impact to  proton transfer in equilibrium conditions   (\cite{Pomes1:1996, Roux:1996, Roux2:2002}). Especially under nonequilibrium conditions, such as the presence of an external electrostatic field   (\cite{Drukker:1998}), or   hydrogen-bonding partners are greatly restricted for the displacements of water molecules in narrow channel pores   (\cite{Decornez:1999}), nuclear tunneling and nonadiabatic transitions need to be accounted in the proton translocation   (\cite{Roux2:2002,Bothma:2010}). Other important works are included in   \cite{SYan:2007, Cukier:2004,Cukier:2004b,Shepherd:2010,Alexov:1999,Cukier:2004,Cukier:2004b,Wang06124516} and  all these studies are based on a full-atom fashion.

	\begin{figure}
    \begin{center}
         \includegraphics[width=1.00\textwidth]{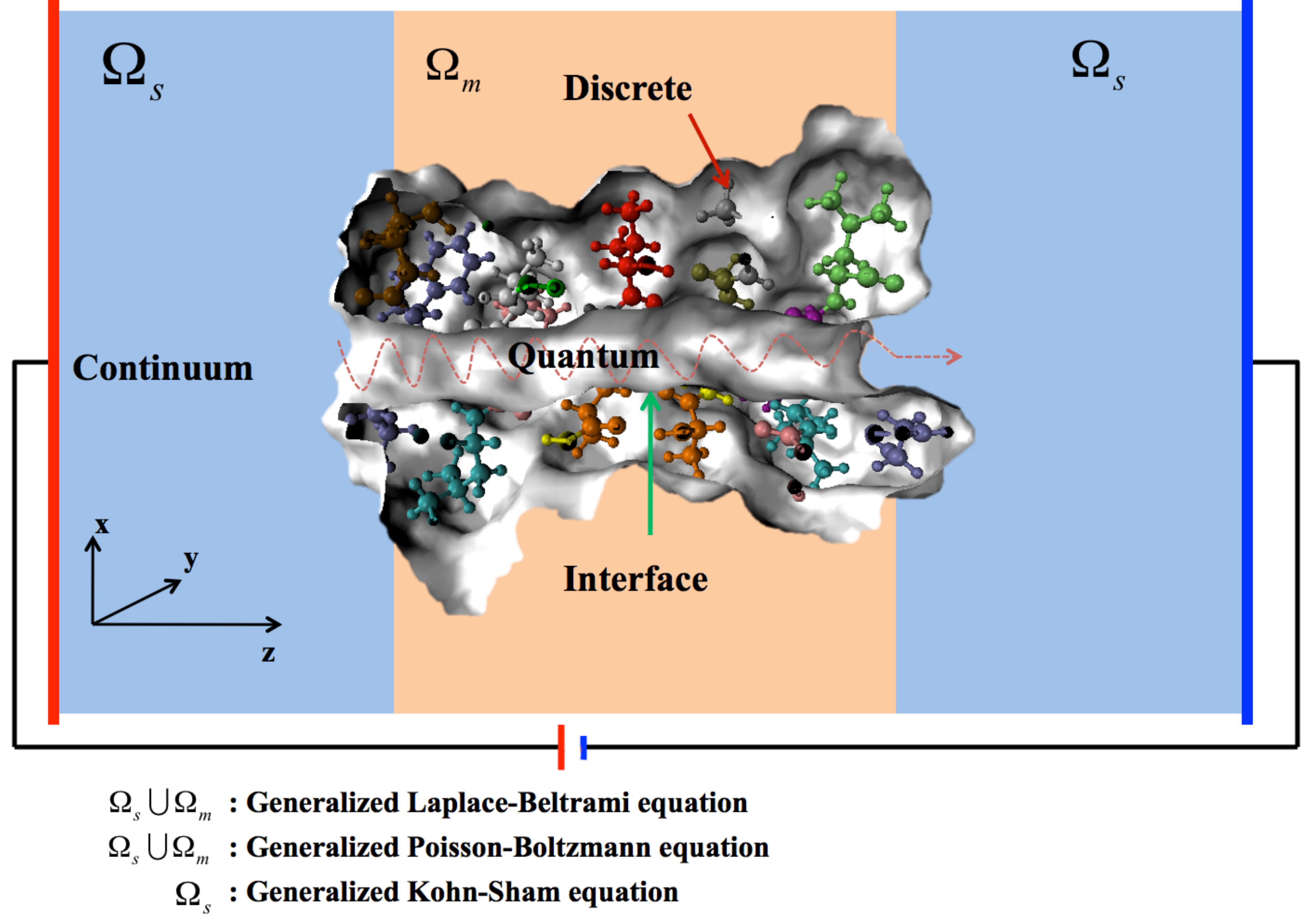}
    \end{center}
    \caption{Illustration of the quantum dynamics in continuum model for proton channels. Dynamics of other ions are modeled by classical Boltzmann approximation in bulk solvent $\Omega_s$, while dynamics of protons is modeled in quantum mechanics but in a mean-field approach.}
    \label{protonmodel}
\end{figure}	
	
		 In recent work by \cite{DuanChen:2013},  a multiscale/multiphysics model was developed for the understanding of the molecular mechanism of proton transport in transmembrane proteins via continuum, atomic and quantum descriptions. As shown in Fig. \ref{protonmodel}, quantum dynamics of proton is expressed in terms of proton concentrations, combined with classical implicit solvent modeling. Further, in order to reduce the number of degrees of freedom,  a new quantum density functional theory was constructed,  based on the Boltzmann statistics to describe proton dynamics quantum mechanically, while numerous solvent molecules  are implicitly treated  as a dielectric continuum. A new density functional like formalism is introduced to represent protein density according to the Boltzmann statistics.   Additionally, generalized correlations that model interactions among  all the ions, and between ions and proteins are explored in detail in  \cite{DuanChen:2011d}. In this model,  proton kinetic and potential energies, the free energy of all other ions, and the polar and nonpolar energies of the whole system are integrated in a multiscale framework on an equal footing. 
 In these models, we consider the energy functional:
\begin{eqnarray}\nonumber
              G_{\rm Total}[\Phi,n, \{\rho_{\beta}\}]&=& \int \left\{- \frac{\epsilon_m}{2}|\nabla\Phi|^2+\Phi \rho_m
                \right.
                \\ \nonumber
                &-&
                \left.
                \frac{\epsilon_s({\bf r})}{2}|\nabla\Phi|^2+\Phi n({\bf r})q  -k_BT\sum_{\beta} \rho_{\beta}^0 \left(e^{-\frac{q_{\beta}\Phi+U_{\beta}-\mu_{\beta 0}}{k_BT}}-1\right)
                \right.
                \\ \nonumber
                &+&
                \left.
                \int\frac{\hbar^2e^{-(E-\mu_p)/k_BT}}{2m({\bf r})} |\nabla\Psi_E({\bf r})|^2dE
                + U_{\rm GC}[n]+ U_{\rm Ext}[n]
                \right.
                \\ \label{eqn:FreeEnergyTotal}
                &+&
                \left.
                 \lambda\left[\frac{N_p}{V_{\Omega}}- \int e^{-\frac{E-\mu_p}{k_BT}} | \Psi_E({\bf r})|^2dE
                \right] \right\}
                d{\bf r}.
        \end{eqnarray}
        This energy functional follows the same pattern of Eq. (\ref{eq17tot}) or (\ref{eq34energy}), but in this case proton is the major ion of interests and its concentration is denoted as $n({\bf r})$. Other ions, with concentration $\rho_{\beta}$, are modeled by the equilibrium Boltzmann distribution as in Eq. (\ref{eq34energy}). 
        
        Proton is described in a quantum mechanical formulation in terms of kinetic and potential energies in the third line of Eq. (\ref{eqn:FreeEnergyTotal}), where $\hbar$ is the reduced Planck constant, and $m(\bf r)$ is the effective mass of the proton. Under this framework, the kinetic energy is represented by the gradient of the proton wavefunction $\Psi_E(\bf r)$. The potential energy of protons includes electrostatic energy $\Phi n({\bf r})q$,  generalized correlation $U_{\rm GC}[n]$, and external $U_{\rm Ext}[n]$, which are approximated as functionals of the proton concentration. 
        The wavefunction and number density of  protons have the following relation:
        \begin{equation}\label{density}
        		n({\bf r})=\int|\Psi_E({\bf r})|^2e^{-\frac{E-\mu_p}{k_BT}}dE,
        \end{equation}
where $\mu_p$ is the general electrochemical potential of protons. 

The total  energy functional (\ref{eqn:FreeEnergyTotal}) represents a multiphysical and multiscale framework that contains the continuum approximation for the solvent, while explicitly takes into account the channel protein in discrete atomic details. More importantly, it puts the classical theory of electrostatics and the quantum mechanical description of protons  on an equal footing. Similar energy frameworks  have been developed for  nano-electronic devices in   \cite{DuanChen:2010}.


The governing equation for the electrostatic potential can be derived by the variation of  energy functional (\ref{eqn:FreeEnergyTotal}) with respect to electrostatic potential  $\Phi$
\begin{equation}\label{eqn:PoissonEq}
    \frac{\delta G_{\rm Total}[\Phi,n]}{\delta \Phi}=0 \Longrightarrow
    -\nabla\cdot\left(\epsilon\nabla\Phi({\bf r})\right)-\sum_{\beta}^{N'_c}\rho^0_{\beta}{e}^{-\frac{q_{\beta}\Phi({\bf r})-\mu_{\beta}}{k_BT}}
    = n({\bf r})q + \rho_f({\bf r}).
\end{equation}

In the present multiphysics  model, the proton number density $n({\bf r})$ in Eq. (\ref{eqn:PoissonEq}) is  related to the wavefunction $\Psi_E({\bf r})$, which is governed by the generalized Kohn-Sham equation. This equation is obtained by the variation of the total free energy functional (\ref{eqn:FreeEnergyTotal}) with respect to wavefunction $\Psi_E^{\dagger}$
\begin{equation}\label{eqn:Kohn-Sham}
\frac{\delta G_{\rm Total}[\Phi,n]}{\delta \Psi_E^{*}} =0\Longrightarrow
    -\nabla\cdot\frac{\hbar^2}{2m({\bf r})}\nabla\Psi_E({\bf r}) + V({\bf r})\Psi_E({\bf r})
  =E\Psi_E({\bf r}),
\end{equation}
where we set the Lagrange multiplier $\lambda=E$. The total Hamiltonian of the proton is given by 
\begin{equation}
    H=-\nabla\cdot\frac{\hbar^2}{2m({\bf r})}\nabla + V({\bf r}),
\end{equation}
 in which the total effective potential energy
 \begin{equation}
    V({\bf r})= q\Phi({\bf r}) + V_{\rm GC}({\bf r})+ V_{\rm Ext}({\bf r})
 \end{equation}
consists of electrostatic, generalized correlation and external contributions. The external potential can be omitted for a closed system without external fields.

It is important to note that generalized Kohn-Sham equation (\ref{eqn:Kohn-Sham}) is fundamentally different from the normal Kohn-Sham equation for electronic structures. The Kohn-Sham operator in Eq. (\ref{eqn:Kohn-Sham}) has an 
absolutely continuous spectrum and invokes the Boltzmann statistics for proton scattering. Whereas the normal Kohn-Sham operator has a discrete spectrum and assumes the Fermi Dirac statistics for electron occupations (bound states). Eqs. (\ref{eqn:PoissonEq}) and (\ref{eqn:Kohn-Sham}) form the Poisson-Boltzmann-Kohn-Sham (PBKS) equations.

From the above quantities, the proton current can be defined by standard probability flux, whose practical expression is the following
\begin{equation}\label{eqn:protoncurrent}
I=\frac{q}{h}{\rm Tr} \int G(E) {V}^{ah}_{\rm intra}G^{\dagger}(E)V^{ah}_{\rm extra}\left[e^{-\frac{E-\mu_{\rm extra}}{k_BT}} -  e^{-\frac{E-\mu_{\rm intra}}{k_BT}}\right]dE.
\end{equation}
where ${\rm Tr}$ is the trace operation, $G$ is the Green's operator
\begin{equation}
	G(E)=(E - H)^{-1},
\end{equation}
and  $\mu_{\rm extra}$ and $\mu_{\rm intra}$ are the external electrical field energies at extracellular  and intracellular electrodes, respectively. Here $V_{\rm extra}^{ah}$ and $V_{\rm intra}^{ah}$ are the anti-Hermitian components of the external potentials   (\cite{DuanChen:2013}).

\begin{figure}[ht!]
	 	\begin{center}
			\begin{tabular}{cc}
                        \includegraphics[width=0.5\textwidth]{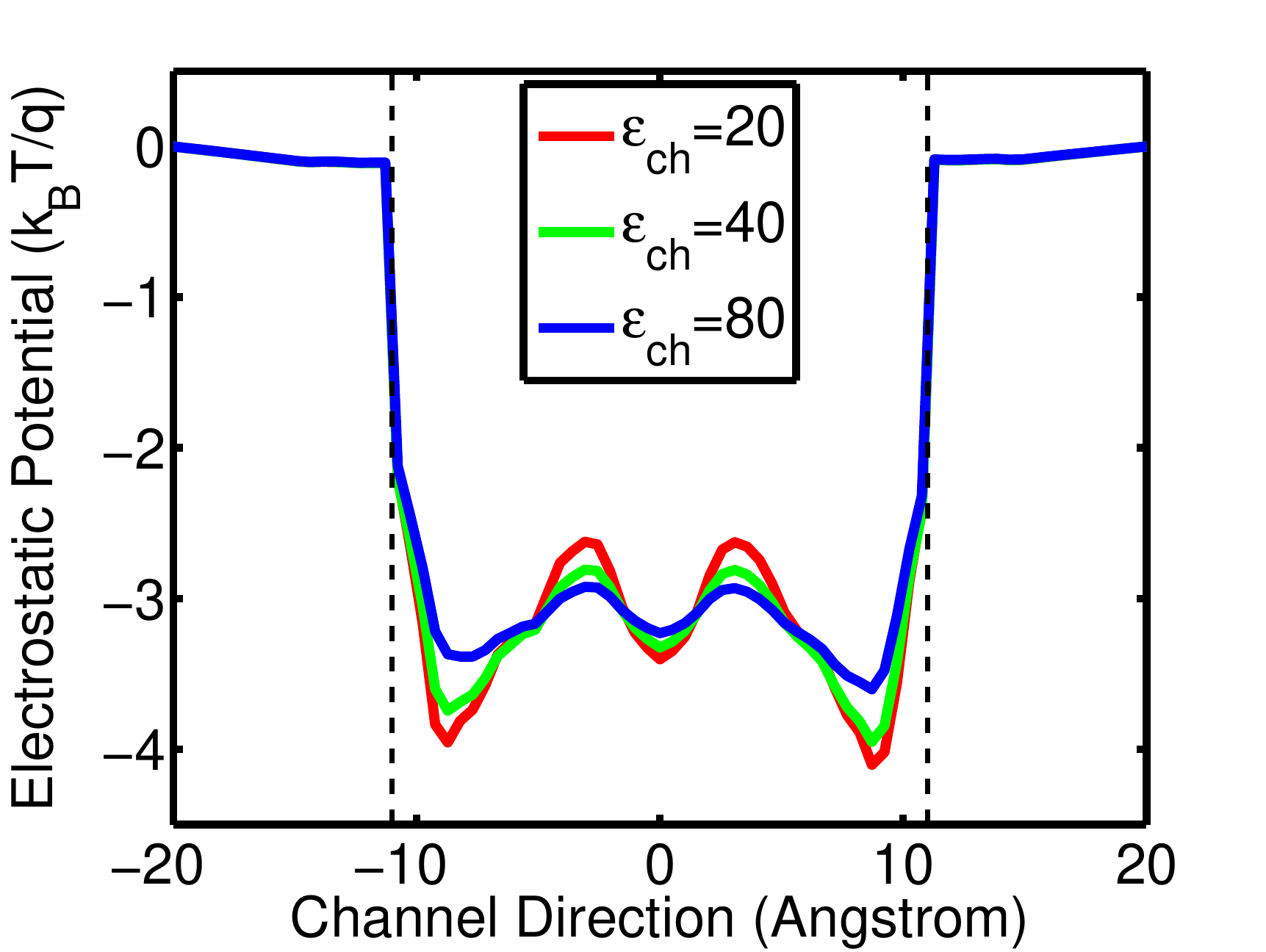}&
                        \includegraphics[width=0.5\textwidth]{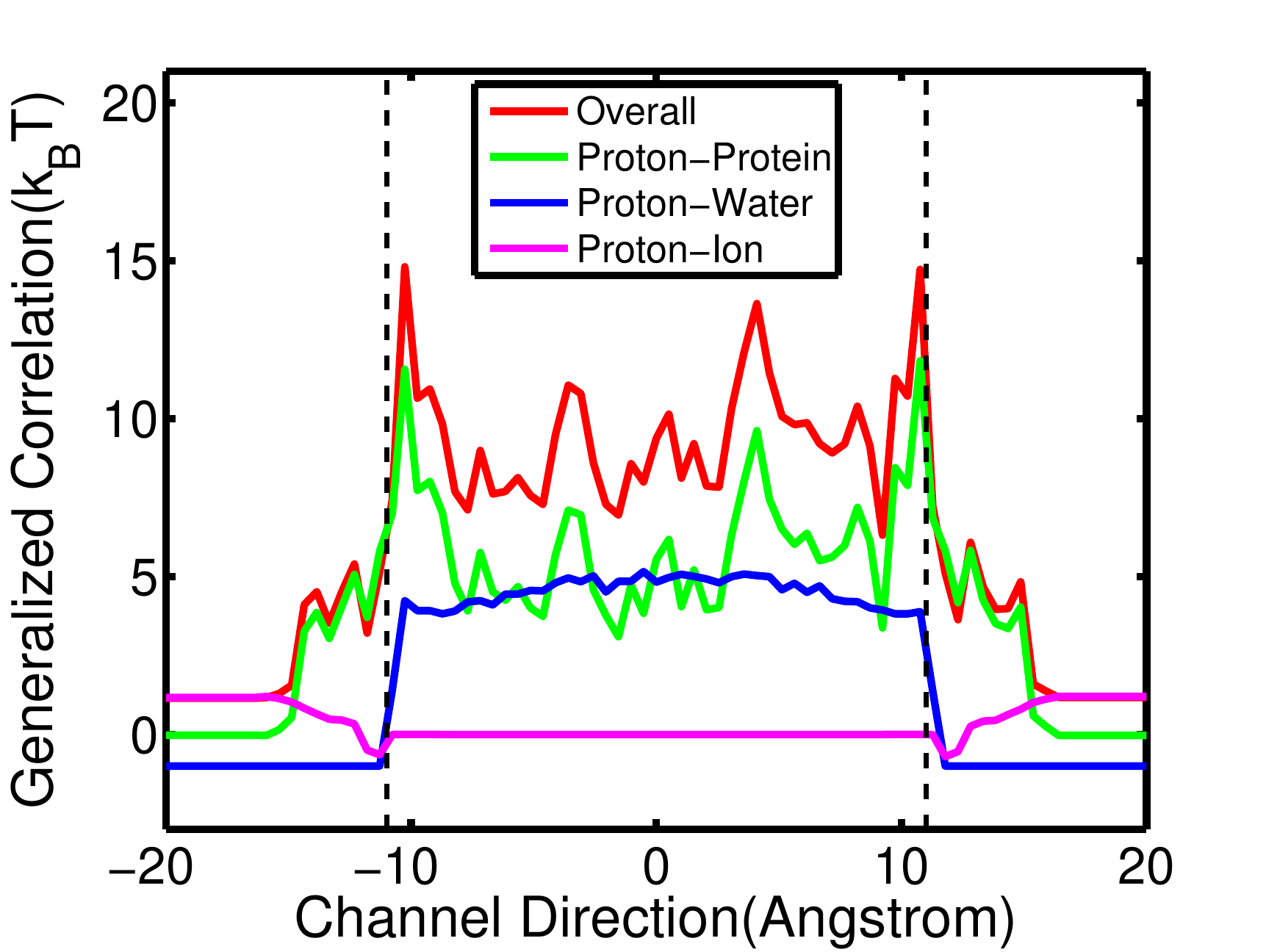}\\
                        (a) & (b)
                        \end{tabular}
		\end{center}
          	\caption{Potential energy components of proton transport through  the GA channel along the $z$-axis. (a) Electrostatic potential energy. (b) Generalized correlations}
            			\label{fig:protonphi}
                \end{figure}

	Figures \ref{fig:protonphi} (a)-(b) display the electrostatic potential energy and the generalized correlations of protons in the GA channel, respectively. The electrostatic potential energy in this model is very similar to the profiles in the PNP or PBNP model, because the electrostatics is determined majorly by the molecular structure of the GA channel. The dielectric constant of water molecules in the channel region is taken as a model parameter and three values, $\epsilon_{\rm ch}=20$, $\epsilon_{\rm ch}=40$,  and $\epsilon_{\rm ch}=80$ are used in Fig. \ref{fig:protonphi} (a). Two electrostatic potential wells present near the entrance and exist of the channel, corresponding  to the two binding sites. In Fig. \ref{fig:protonphi}(b) there shows the generalized correlation of protons, including protein-proton, proton-water, and proton-ion interactions, modeled by density functionals. In this case, generalized correlations are all energy barriers that protons need to overcome during the transport, and the sum of electrostatics and generalized correlation is the overall potential energy of protons.
                
\begin{figure}[ht!]
         \begin{center}
         		\begin{tabular}{cc}
         			  \includegraphics[width=0.5\textwidth]{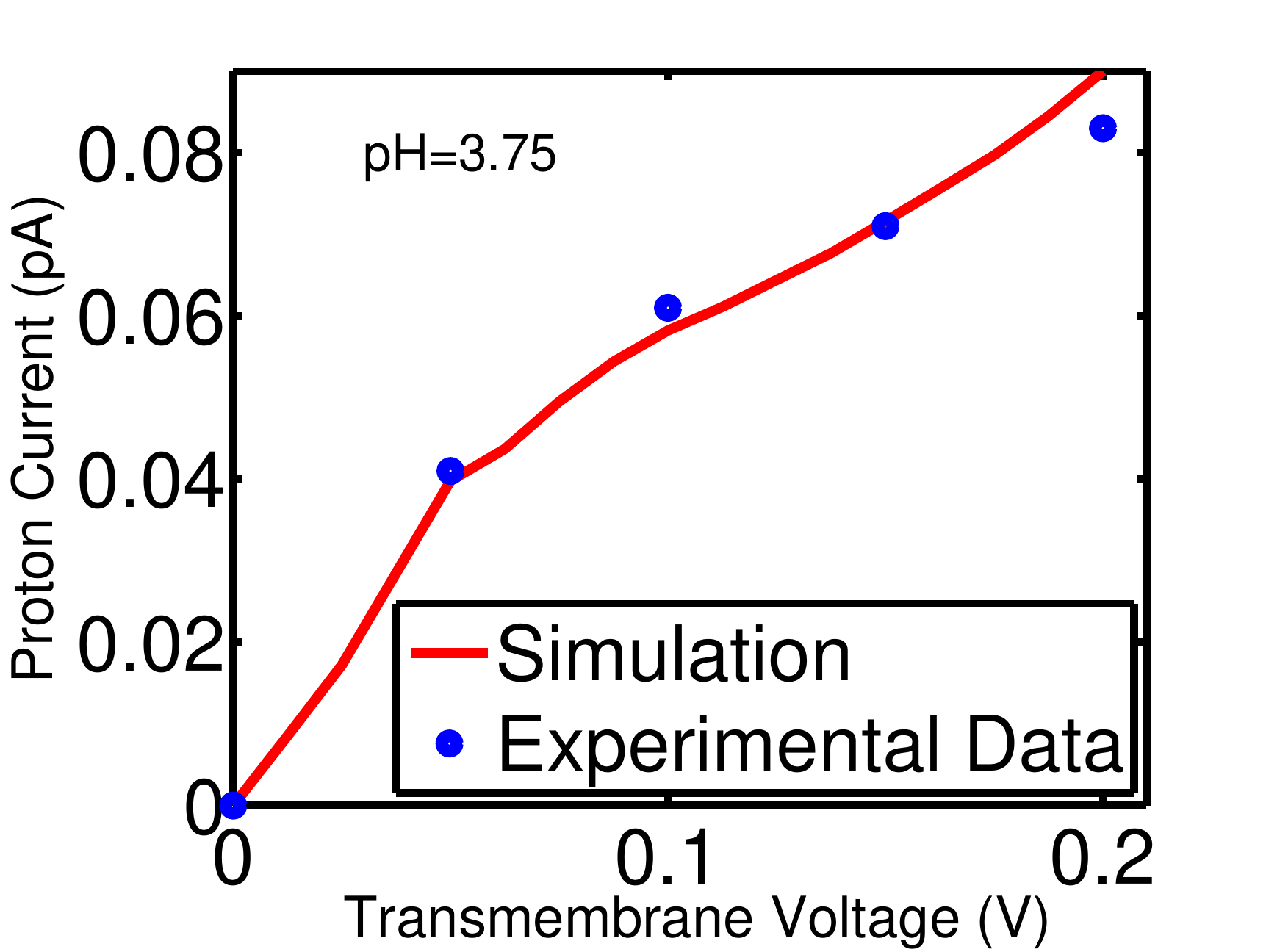}&
          		 \includegraphics[width=0.5\textwidth]{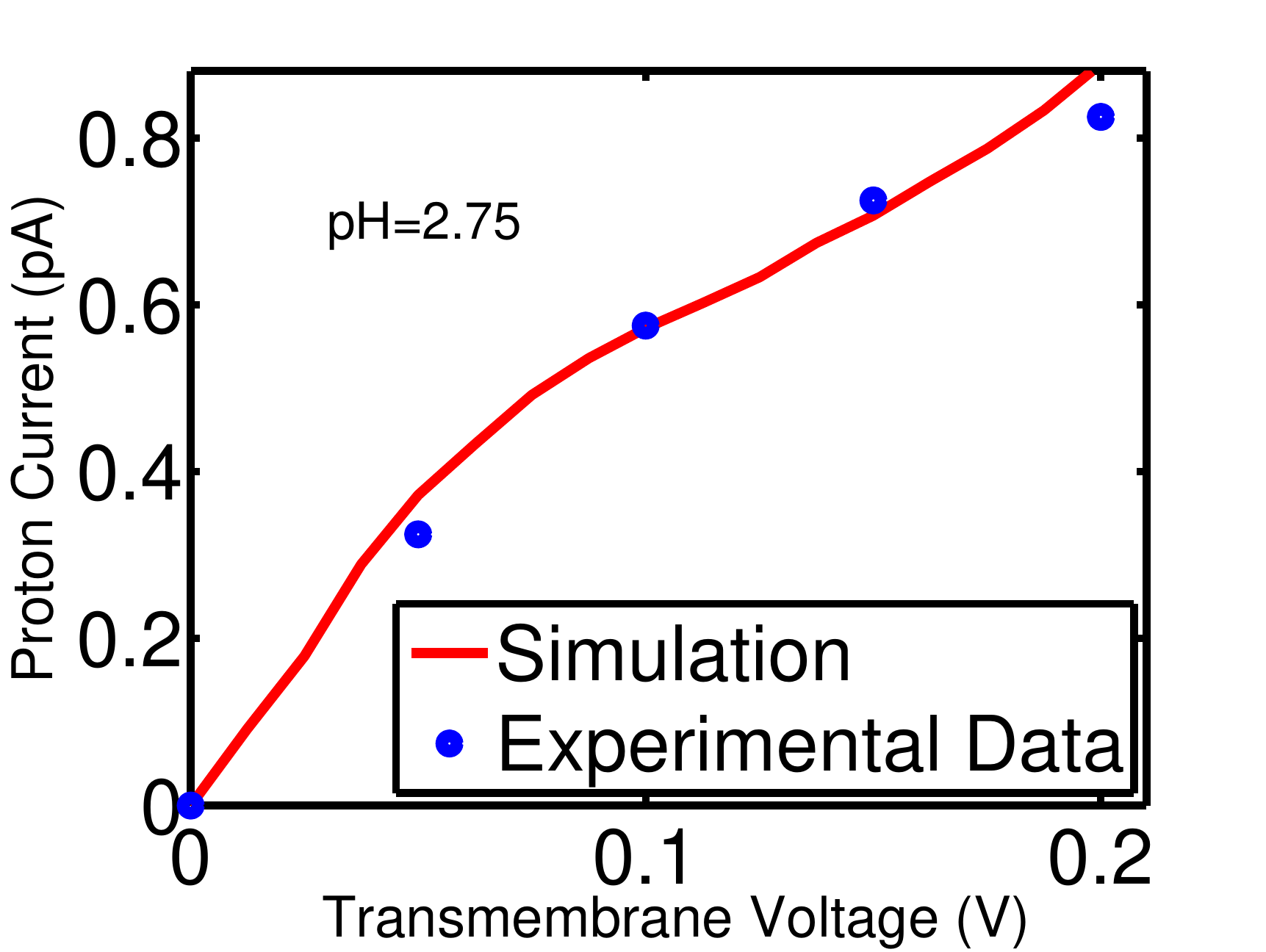}\\
				(a) pH=3.75 & (b) pH=2.75
           	\end{tabular}
          \end{center}
          \caption{  Voltage-current relation of proton translocation of GA at different concentrations.
          		Blue dots: experimental data of \cite{Eisenman:1980}; Red solid curves: QDC model prediction.
          			  }
            			\label{ivgc}
	\end{figure}

	Figure \ref{ivgc} displays simulated results of proton conductance through Eq. (\ref{eqn:protoncurrent}), compared with  experimental data from \cite{Schumaker:2000} of the GA channel. The blue dots in each figure represent available experimental observations for certain voltage biases, while the red curves are from the PBKS model predictions calculated with sufficiently many voltage samples. The model parameters  are chosen to match the experimental data but all of the choices are taken within the range of physical measurements. 

	\subsection{Differential geometry (DG) based charge transport models}\label{sec:DG}

	Like other biological processes, charge transport takes place in an aqueous environment because 65\%-90\% of cell mass is water. Thus, understanding the solvation process of channel proteins and mobile ions in solvent has equal importance as ionic dynamics. In terms of modeling, a solvation process may include the creation of a solute cavity in the solvent, the hydrogen bond breaking and formation at the solvent-solute interface, the surface reconstruction of the solute molecule, and/or the entropy effect due to solvent-solute mixing (\cite{Mukhopadhyay:2012,Onufriev:2002}). Physically, this process involves a variety of solvent-solute interactions, such as the electrostatic, dipolar, induced dipolar, and van der Waals interactions between the solvent and the solute (\cite{Ren:2013,Rocchia:2001,Rocchia:2002,Baker:2003,Baker:2006}).
		In the past several years, intensive investigations for modeling ion channel systems based on the PNP theory but with solvation processes through multiscale, multiphysics variational approach were carried out, which were based on a differential geometry based multiscale paradigm  for large chemical and biological systems, such as fuel cells, nanofluidics (\cite{JinPark:2015}), ion channels, molecular motors, and viruses   (\cite{Wei:2009,DuanChen:2013,DuanChen:2011c,DuanChen:2011d,ZhanChen:2010a,ZhanChen:2010b,ZhanChen:2011a,Wei:2012,Wei:2013}). With abundant water molecules and atomic details in cell membrane,  it is critical to perform dimensionality  reduction and manifold contraction by multiscale approaches. The essential ingredient  is to use the differential geometry theory and analysis for surfaces and geometric measure  as a natural technique to distinguish macroscopic domain for solvent and membrane, from the microscopic domain for channel proteins. At the same time, the differential geometry based model couples the continuum mechanical description of the aquatic environment with the discrete atomistic description of ion channels.

		These differential geometry based multiscale models are  intensively investigated  and practically implemented for various types of ion channel problems    (\cite{ZhanChen:2010a,ZhanChen:2010b,ZhanChen:2011a,DuanChen:2011c,DuanChen:2011d}).  There are two representations,  the Eulerian formulation   (\cite{ZhanChen:2010a, ZhanChen:2011a, ZhanChen:2012, BaoWang:2015a}) and the Lagrangian formulation   (\cite{ZhanChen:2010b}) for the key element in the models, the solvent-solute interface (\cite{Bates:2008,Bates:2009}). 
For the former, the interface is  described  as a hypersurface function which is evolved according to the derived governing equations   (\cite{ZhanChen:2010a}), while for the latter,  interface elements are directly evolved according to  governing equations which prescribe a set of rules.
The Lagrangian representation of a molecular surface can be obtained from the projection of the hypersurface function by using an isosurface extraction procedure. The Eulerian formulation  is mathematically simple and computationally robust,  while the Lagrangian formalism is straightforward for force prescription   (\cite{Bates:2009}) and is computationally efficient, but usually encounters difficulties in handling the geometric break-up and/or surface merging. Validation and equivalence of these two formulations are tested by  the solvation analysis  for biological and chemical compounds  (\cite{ZhanChen:2010a, ZhanChen:2010b}). These models were shown to deliver excellent solvation predictions of experimental data (\cite{ZhanChen:2012, BaoWang:2015a}).

		\paragraph{Variational solute-solvent interface}
		
			The first ingredient of the DG model is the definition of molecular surface. Implicit solvent models  require  a given solvent-solute interface, or molecular surface,  to distinguish different domains with the corresponding physical features, e.g.,  dielectric functions and diffusion constants,  and to separate appropriate computational domains.   In many models, simple {\it ad hoc} molecular surfaces, such as the van der Waals surface, the solvent excluded surface   (\cite{Richards:1977}), or the solvent accessible surface are often utilized in applications of protein-protein interactions   (\cite{Crowley}),  protein folding   (\cite{Spolar}), DNA binding and bending   (\cite{Dragan}).  In \cite{Wei:2005}, the first PDE-based approach to construct  biomolecular surfaces via curvature driven geometric flows was  introduced. Later on, the first variational formulation of molecular surfaces is developed, and the resulting molecular surface, called the minimal molecular surface (MMS), was constructed by the mean curvature flow   (\cite{Bates:2006,Bates:2008}). Physically, the new definition of molecular surface satisfies the physical requirement of free energy minimization of the ion channel system. Computationally, it avoids artificially geometric defects such as cusps or self-intersecting surfaces, which could lead to computational instabilities.

\begin{figure}
            \begin{center}
                         \includegraphics[width=0.6\columnwidth]{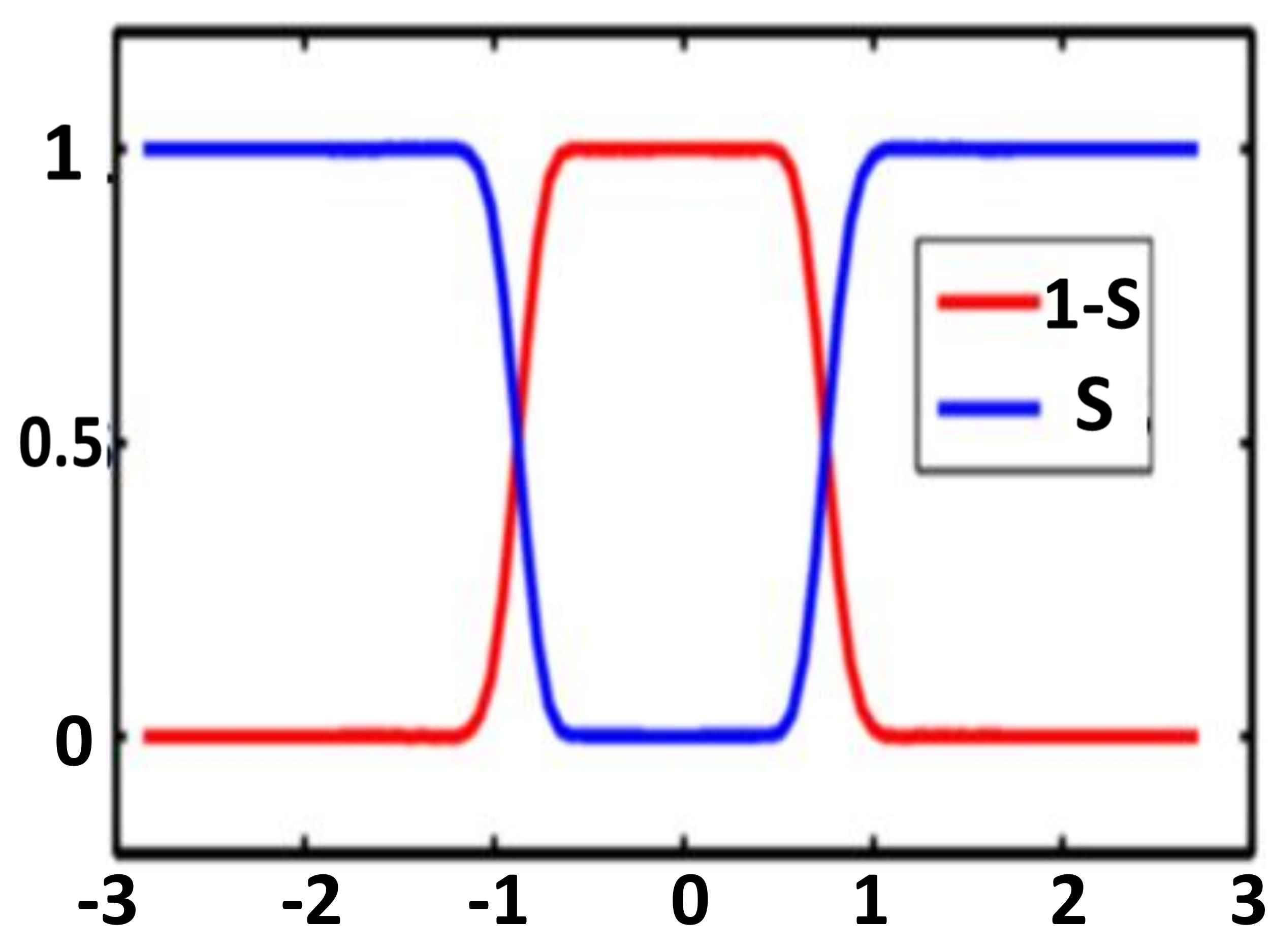}
            \end{center}
             \caption{Illustration of surface  characteristic function $S$ and solvent characteristic function $1-S$ in a 1D setting. }
             \label{figS-profile}
\end{figure}

			To develop the variational solute-solvent interface, a solute characteristic function $S({\bf r})$ is introduced in the total free energy. Fig. \ref{figS-profile} offers a 1D representation, in which $S({\bf r})$ takes values one and zero in the solute and solvent domains, respectively but with a transient region. Correspondingly, the solvent characteristic  function, $1-S({\bf r})$, represents the solvent domain. With this setup, the energy in Eq. (\ref{eq17tot}) can be rewritten as

\begin{eqnarray}
\begin{aligned}
G^{\rm PNP}_{\rm{total}}&[S,\Phi,\{\rho_\alpha\}] = \int \left\{ \gamma |\nabla S | +   p S +   (1-S)U \right.  \\ 
& \left.
+S\left[  -\frac{\epsilon_m}{2}|\nabla\Phi|^2 + \Phi \ \rho_m\right] +(1-S)\left[-\frac{\epsilon_s}{2}|\nabla\Phi|^2
+\Phi\sum_{\alpha} \rho_{\alpha}q_{\alpha} \right]\right.  \\\label{geq17tot}
& \left. +(1-S)\sum_\alpha
\left[   \left(\mu^0_{\alpha }-\mu_{\alpha 0} \right) \rho_\alpha + k_B T  \rho_\alpha  {\rm{ln}} \ \frac{\rho_\alpha }{ \rho_{\alpha 0} } - k_B T \left(\rho_\alpha  - \rho_{\alpha 0} \right)  + \lambda_\alpha \rho_\alpha \right]
\right\}
d{\bf{r}}.
\end{aligned}
\end{eqnarray}
The first row of Eq. (\ref{geq17tot}) lists the nonpolar solvation free energy of the system, with $\gamma$ being the surface tension and $p$ being the hydrodynamic pressure, respectively. The solvent-solute interaction $U$  is originally 
approximated by     \cite{ZhanChen:2010a,ZhanChen:2010b},
the Lennard-Jones potential, i.e.,
\begin{equation}\label{eqn:olddef}
	U = \sum_j^{Na}U_j^{LJ}({\bf r})
\end{equation}
where $Na$ atoms are assumed for the channel protein. The Weeks-Chandler-Andersen (WCA) decomposition based on the original WCA theory from  \cite{Weeks:1971} is utilized to split the Lennard-Jones potential into attractive and repulsive parts. 
The WCA potential was used to account for the attractive dispersion interaction  \cite{ZhanChen:2010a,ZhanChen:2010b}, but can be extended as the generalized correlation in terms  of ionic concentrations, i.e.,
\begin{equation}
	U=\sum_{\alpha}\rho_{\alpha}U_{\alpha}.
\end{equation}
Details of the formation and explanation of the generalized correlation $U_{\alpha}$ is given in the later Section.

The second and third rows of Eq. (\ref{geq17tot}) follows the same concepts in Eq. (\ref{eq17tot}), except that the solvent and solute characteristic functions are associated correspondingly.


\paragraph{Generalized Laplace-Beltrami (LB) equation}
By applying the  variational principle, the governing equation for the function $S({\bf r})$ is 
\begin{eqnarray}  \label{eq18vars}
\begin{aligned}
\frac{\delta G^{\rm PNP}_{\rm{total}}}{\delta S}\Rightarrow & -\nabla\cdot\left(\gamma\frac{\nabla S}{|\nabla S |}\right)
   + p -  U
                -\frac{\epsilon_m}{2}|\nabla\Phi|^2 + \Phi \ \rho_m \\
                &+\frac{\epsilon_s}{2}|\nabla\Phi|^2 -\Phi\sum_{\alpha} \rho_{\alpha}q_{\alpha}
               -\sum_\alpha \left[ -\mu_{\alpha 0}  \rho_\alpha  + k_B T  \rho_\alpha  {\rm{ln}} \ \frac{\rho_\alpha }{ \rho_{\alpha 0} } - k_B T \left(\rho_\alpha  - \rho_{\alpha 0} \right)   \right]=0 ,
\end{aligned}
\end{eqnarray}
where  Eq. (\ref{eq20Equil}) is used in the derivation. It is easier to pursue the solution of Eq. (\ref{eq18vars}) by the following parabolic equation with an artificial time  (\cite{Bates:2009,Wei:2009,SZhao:2011a}):
\begin{eqnarray}\label{eq25surf}
   \frac{\partial S}{\partial t}&=&|\nabla S|\left[\nabla\cdot\left(\gamma\frac{\nabla S}{|\nabla S|}\right)
   + V_1\right],
\end{eqnarray}
where the LB potential $V_1$ is
\begin{eqnarray}\label{eq26vterm}
   V_1&= &- p + U
                +\frac{\epsilon_m}{2}|\nabla\Phi|^2 - \Phi \ \rho_m-\frac{\epsilon_s}{2}|\nabla\Phi|^2+\Phi\sum_{\alpha} \rho_{\alpha}q_{\alpha}  \\ \nonumber
                &&+\sum_\alpha  \left[k_B T  \left( \rho_\alpha  {\rm{ln}} \ \frac{\rho_\alpha }{ \rho_{\alpha 0} } -\rho_\alpha  +\rho_{\alpha 0} \right)  -\mu_{\alpha 0}  \rho_\alpha \right].
\end{eqnarray}

\begin{figure}[ht!]
    \begin{center}
           \begin{tabular}{cc}
         	  \includegraphics[width=0.5\textwidth]{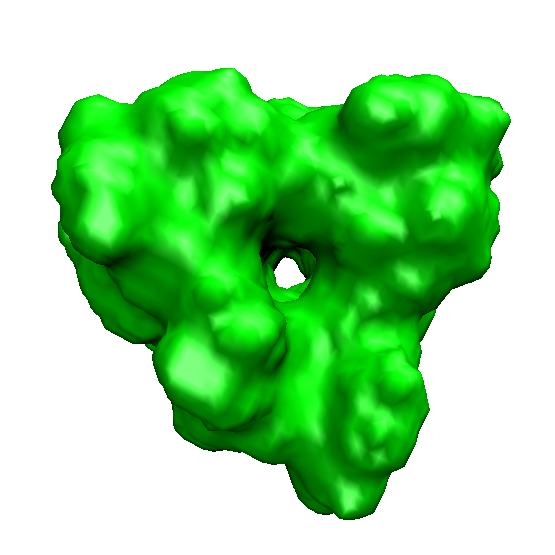}&
          	  \includegraphics[width=0.5\textwidth]{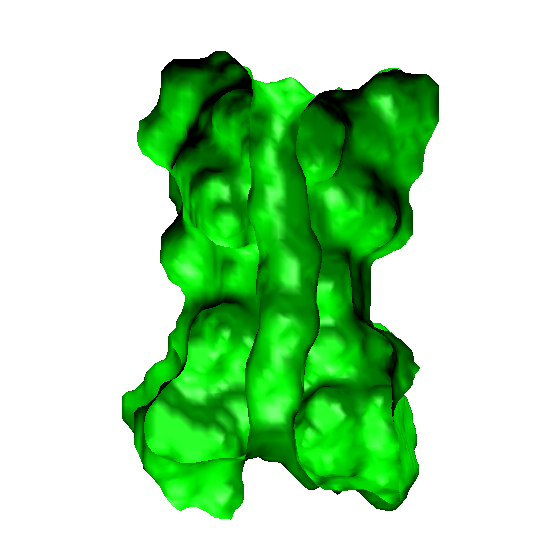}\\
		(a)  & (b) \\
                \includegraphics[width=0.5\textwidth]{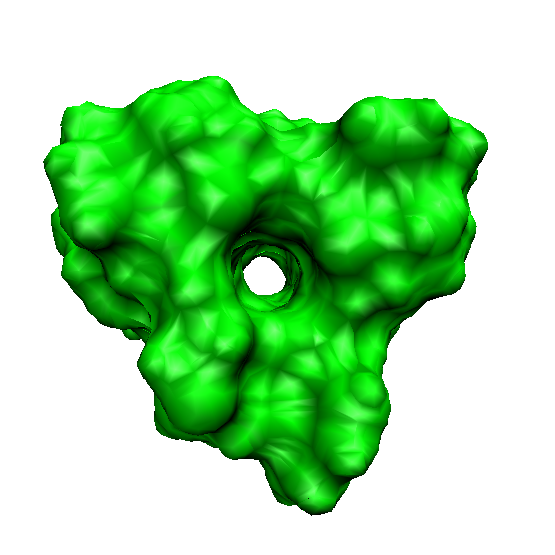}&
          	  \includegraphics[width=0.5\textwidth]{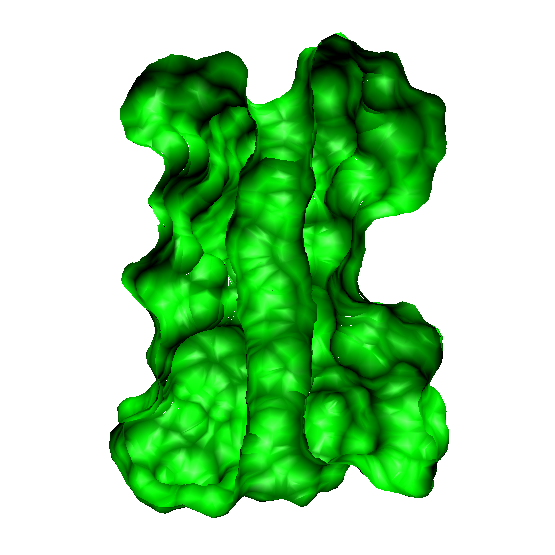}\\
		(c)  & (d) \\
           \end{tabular}
    \end{center}
       \caption{Surface representations of the GA channel.  (a)-(b) Surface extracted from the generalized LB equation with $S=0.5$. (c)-(d) MSMS surface with probe radius 1.4 and density 10.}
    \label{fig:vs-ses}
\end{figure}

	Figures \ref{fig:vs-ses} (a)-(b) show the molecular surface generated from the generalized LB equation (\ref{eq25surf}). For comparison,  the solvent exclusive surface (SES) of the GA generated via the MSMS package  developed by \cite{Sanner:1996} is also displayed in Figs. \ref{fig:vs-ses} (c)-(d).  The SES only depends on the parameter used in the generating software package (water probe radius 1.4 \AA~ and density 10). Once generated, it keeps fixed and is independent of the physiological conditions such as ion concentration or transmembrane voltage differences in simulations. On the other hand,  generation of the surface  from the generalized LB equation is an iteration process as stated in the earlier sections. As included in the LB potential $V_{1}(\Phi, \{\rho_{\alpha}\})$, evaluation of the characteristic functions $S(\bf r)$ depends on the electrostatics and ionic concentration. The results showed in Figs. \ref{fig:vs-ses}(a) and (b) are calculated under the transmembrane potential of 0.2 mV and with the ionic concentration of 0.1M.

    Comparing Figs. \ref{fig:vs-ses}(a) and (b) and Figs. \ref{fig:vs-ses}(c) and (d), the SES commits geometric singularities, such as cusps and self-intersecting surfaces  \cite{QZheng:2011a}. These singularities may bring computational difficulties to the designed interface schemes and are unphysical in the solvent-solute interface. In contrast, the surface from the LB equation has a smoother appearance because of the diffusion mechanism, which gives less-intensive sharp changes near the solvent-solute boundary. More importantly, generations of the new molecular surface incorporate the interactions with the external ionic condition as well as transmembrane voltages. All of these characteristics make more physical sense, and many good results in applications are obtained  in \cite{ZhanChen:2010a,ZhanChen:2010b}.

%
%

	\paragraph{Generalized Poisson-Nernst-Planck equation}
Under this differential geometry based model,  the generalized Poisson-Nernst-Planck equation can be derived   by taking the variation with respect to the electrostatic potential $\Phi$, i.e, 
\begin{eqnarray}\label{geq24poisson}
-\nabla\cdot\left(\epsilon(S) \nabla\Phi \right)= S\rho_m
    +(1-S)\sum_{\alpha} \rho_{\alpha}q_{\alpha},
\end{eqnarray}
where $\epsilon(S)=(1-S)\epsilon_s+S\epsilon_m$ is an interface-dependent dielectric function. Eq. (\ref{geq24poisson}) depends on  ion concentration  $\rho_{\alpha}$ and  the solute characteristic function $S$.

\begin{eqnarray}\label{geq22nernst}
 \frac{\partial \rho_\alpha}{\partial t}=\nabla \cdot \left[D_{\alpha}
  \left(\nabla \rho_{\alpha}+\frac{ \rho_{\alpha}}{k_{B}T}\nabla (q_\alpha\Phi +U_{\alpha})\right)\right],
\end{eqnarray}
This is the  generalized Nernst-Planck equation where $q_\alpha\Phi +U_{\alpha}$ is a mean-field  approach of all potentials.  Equation (\ref{geq22nernst}) reduces to the standard Nernst-Planck equation in Eq. (\ref{eq22nernst}) when the solvent-solute interactions are taken as the original definition in Eq. (\ref{eqn:olddef}).
The differential geometry (DG) based model can also be applied to the PBNP and PBKS systems similarly. 

		\paragraph{Generalized correlation}\label{sec:GC}

	Many improvements for the PNP system, such as the size effects or ion-water interactions discussed earlier, can be adopted in the differential geometry based model, termed as generalized correlation, which are modeled in a mean-field approach. 

 In this approach,  the generalized correlation is modeled by extending  the term
\begin{equation}\label{eqn:gc}
	\int (1-S)Ud{\bf r}
\end{equation}
in Eq. (\ref{geq17tot}) as
the free energy functional of  the local ionic concentration $\rho_{\alpha}$  and its gradient $\nabla \rho_{\alpha}$ for all ions,  i.e., 
\begin{equation} \label{Eqn:nonelec}
G_{{\rm GC}}[S,\{\rho_{\alpha}\}]=\int (1-S({\bf r}))\sum_{\alpha}U_{\rm GC}[\{\rho_{\alpha}\},\{\nabla \rho_{\alpha}\}]d{\bf r}.
\end{equation}

With the assumption that the  $\nabla \rho_{\alpha}$ dependence is omitted as a first order approximation, it takes
 \begin{eqnarray}\label{eqn:GC1}
	G_{{\rm GC}}[S,\{\rho_{\alpha}\}]&=&\sum_{\beta}\left(\frac{1}{2}\right)^{\delta_{\alpha\beta}}\int\int (1-S({\bf r}))\rho_{\alpha}({\bf r})\rho_{\beta}({\bf r}')K_{\beta}(|{\bf r}-{\bf r}'|)d{\bf r}'d{\bf r}\\\label{eqn:GC2}
			&+& \int\int (1-S({\bf r})) \rho_{\alpha}({\bf r})n_w({\bf r}')K_w(|{\bf r}-{\bf r}'|)d{\bf r}'d{\bf r}\\\label{eqn:GC4}
			&+&\sum_{j=1}^{N_a}\int (1-S({\bf r})) \rho_{\alpha}({\bf r})n_j({\bf r}')K_j(|{\bf r}-{\bf r}'|)d{\bf r},
\end{eqnarray}
where $\delta_{\alpha\beta}$ is the Kronecker delta function with $\delta_{\alpha\beta}=1$ if $\alpha=\beta$ and $\delta_{\alpha\beta}=0$ when $\alpha\neq\beta$. The kernels $K_{\beta}(|{\bf r}-{\bf r}'|)$, $K_w(|{\bf r}-{\bf r}'|)$, and $K_i(|{\bf r}-{\bf r}'|)$ model interactions among  the $\alpha$-th and $\beta$-th ions, water molecules (in sense of density $n_w({\bf r})$), and protein atom distribution $n_i({\bf r}')$, respectively.

Finally, by taking variation with respect to the concentration $\rho_{\alpha}$, one obtains the generalize correlation potential 
\begin{equation} \label{Eqn:nonelec2}
U_{\alpha}({\bf r})=\sum_{\beta}\rho_{\beta}({\bf r})\ast K_{\beta}({\bf r})+n_w({\bf r})\ast K_w({\bf r})+\displaystyle\sum_{j=1}^{N_a}n_j({\bf r})\ast K_j({\bf r}),
\end{equation}
where $*$ represents the convolution operation.



\section{Concluding remarks}


Phenomena of charge transport  present not only  in naturally-designed devices such as ion channels, deoxyribonucleic acid (DNA) nanowires, ATPases, or neuron synapses, but also widely exist in human-made devices including solar cells, fuel cells, battery cells, molecular switches, nanotubes, field effect transistors, nanofibers, thin films, etc (\cite{Wei:2012}). Thus, understanding of mechanism and dynamics of charge transport in these nano-scale biological or industrial devices is prerequisite to study their functions for human health or developing modern technologies.  Quantitative modeling and simulation of charge transport have emerged as a new field in applied mathematics, which offers 
mathematical models, algorithms and analysis to reproduce experimental data, to predict new phenomena and hence  to shed light on new directions of research for these nano-bio systems. Due to  complexity of charge transport increases dramatically at nanoscale, where a large amount of components, both macroscopic and microscopic, interact with each other in multiphysical principles in a heterogeneous environment,  Mathematical modeling and computation of these complex systems encounter formidable challenges.  
 Fully atomic description for the whole system will offer the highest accuracy, but it is prohibited by the intractable  large number of degree of freedom. Multiscale and multiphysics modeling that retains   the atomistic description of the channel protein while treats the membrane and solvent as dielectric continuum gives rise to efficient approaches to ion channel charge transport. 

In this paper we review recent progresses in mathematical modeling, algorithm  and analysis  of charge transport in ion channels. A major emphasis is the development in Poisson-Nernst-Planck (PNP) based models. The PNP equations can be derived from the total energy functional of an ion channel system and include a Poisson equation for electrostatic environment for the whole system and a series of Nernst-Planck equations for dynamics of mobile ions. Modified PNP models were developed under this framework in order to include more detailed physical properties such as ion size effects or ion-water interactions, or to reduce model complexity in multi-ion species systems. All the modification and improvement of PNP equations can be achieved by adding additional energy components in the total energy functional. On the other side, proton transport is a special type of charge transport due to the properties of hydrogen ions, and thus quantum dynamics is involved in modeling proton channels. There are quite a few literature about quantum models for proton transport in the literature, but it is the first time that a quantum dynamics in continuum model was established as a mean-field method. Finally, a differential geometry (DG) based  multiscale model was developed, which includes the solvation process of ion channels in a solvent. In the multiscale treatment of the ion channel system, the DGPNP model generates a more reliable and robust definition of molecular surface, which takes into account the mutual interactions with electrostatics and ionic dynamics. Various computational algorithms and mathematical analysis are also reviewed in this work.

	Although numerous efforts and work have been devoted in this area each year worldwide, including the availability of molecular structures of new ion channels, resolving a complete picture of molecular mechanism of channel gating and charge transport  remains a challenging task.
On the one hand, structure determination of ion channel proteins and	in general, G-protein-coupled receptors (GPCRs), is typically more difficult than that of global proteins. On the other hand, even with available structures,  biophysical understanding of functioning principles for many existing ion channels are not completely clear. In particular, the gating mechanism of voltage gated eukaryotic sodium channels is still elusive.     

	
Future mathematical modeling and simulation of ion channels will address the pressing needs in the understanding of their molecular mechanism. Much attention will be paid to the homology modeling of ion channel structures, such as the structure presented in Fig. \ref{fig:nachannel} and their improvement.  Additionally,  the interactions of ion channels and drugs, which hold the key  for the drug discovery for epilepsy, irregular cardiac arrhythmias, hyperalgesia, myotonia, and anesthesia, will be studied by molecular docking. An important component in docking analysis is the prediction of protein-ligand/drug binding affinities, which can be achieved by machine learning algorithms \cite{BaoWang:2016FFTB}.   Moreover, the blind prediction of mutation impacts to ion channel current-voltage (I-V) curves  (\cite{ZXCang:2016b}) will be a new topic in mathematical molecular bioscience and biophysics (\cite{Wei:2016}). In such predictions, machine learning methods can be empowered with mathematical features from differential geometry, algebraic topology, graph theory and partial differential equations (\cite{BaoWang:2016FFTB,ZXCang:2016b}).  Finally, the PNP type of models will be coupled with molecular mechanism and chemical kinetics to address the conformational changes and protonation (or deprotonation) during the ion permeation. Given the importance of the physical and chemical phenomena of charge transport to  both biological systems and nano-device engineering,  theoretical modeling, numerical algorithms, mathematical analysis, and realistic applications of ion channel charge transport will be a focus of  mathematical research in the future.


\section*{Acknowledgments}

The work of GWW was supported in part by NSF Grant   IIS- 1302285 and MSU Center for Mathematical Molecular Biosciences Initiative. The work of the author DC is funded by the Faculty Research Grant 2015-2017 provided by the University of North Carolina at Charlotte.

\newpage
\section*{Literature cited}
\renewcommand\refname{}

\bibliographystyle{agsm}
\bibliography{refs}


\end{document}